%% file: letter.tex
\documentclass[pdftex,twocolumn,epjc3]{svjour3}

\pdfoutput=1

\RequirePackage[T1]{fontenc}

\smartqed  

\RequirePackage{graphicx}
\RequirePackage{mathptmx}      
\RequirePackage{flushend}
\RequirePackage[numbers,sort&compress]{natbib}

\journalname{Eur. Phys. J. C}

\usepackage{preprintcover}
\PreprintCoverPaperTitle{
ATLAS search for a heavy gauge boson decaying to a charged lepton and a neutrino
in {\em pp} collisions at $\sqrt{\pmb s}$ = 7 \tev
}  
\PreprintIdNumber{CERN-PH-EP-2012-245}
\PreprintCoverAbstract{
The ATLAS detector at the LHC is used to search for high-mass states, such as heavy charged gauge bosons (\wp),
decaying to a charged lepton (electron or muon) and a neutrino.
Results are presented based on the analysis of \pp\ collisions at a centre-of-mass energy of
7~\tev\ corresponding to an integrated luminosity of 4.7~\ifb.
No excess beyond Standard Model expectations is observed.
A \wp\ with Sequential Standard Model couplings is excluded at the 95\% credibility level for masses up to 2.55~\tev.
Excited chiral bosons (\wstar) with equivalent coupling strength are excluded for masses up to 2.42~\tev.
}
\PreprintJournalName{Eur. Phys. J. C}

\usepackage{atlasphysics} 
\usepackage{lmetdefs} 
\usepackage{microtype}  
\usepackage{bm}  
\usepackage {multirow}

\begin{document}

\title{ATLAS search for a heavy gauge boson decaying to a charged lepton and a neutrino
in {\em pp} collisions at $\sqrt{\pmb s}$ = 7 \tev}

\author{The ATLAS Collaboration}

\institute{CERN}

\date{Received: date / Accepted: date}

\maketitle

\begin{abstract}
The ATLAS detector at the LHC is used to search for high-mass states, such as heavy charged gauge bosons (\wp),
decaying to a charged lepton (electron or muon) and a neutrino.
Results are presented based on the analysis of \pp\ collisions at a centre-of-mass energy of
7~\tev\ corresponding to an integrated luminosity of 4.7~\ifb.
No excess beyond Standard Model expectations is observed.
A \wp\ with Sequential Standard Model couplings is excluded at the 95\% credibility level for masses up to 2.55~\tev.
Excited chiral bosons (\wstar) with equivalent coupling strength are excluded for masses up to 2.42~\tev.
\end{abstract}

\section{Introduction}
\label{sec:intro}

High-energy collisions at the CERN Large Hadron Collider provide the opportunity
to search unexplored regions for physics beyond the Standard Model (SM) of strong
and electroweak interactions.
One extension common to many models is the existence of additional heavy gauge
bosons, the charged ones commonly denoted \wp.
Such particles are most easily searched for in their decay to a charged lepton (electron or muon)
and a neutrino.

This letter describes such a search performed using
7~\tev\ \pp\ collision data collected with the ATLAS detector during 2011 corresponding to a total
integrated luminosity of 4.7~\ifb.
The data
are used to extend current
limits~\cite{cdf:Wprime2010,cms:wprime2011-2,atlas:wprime_2010_pub,atlas:wprime_2011-2_pub}
on \xbr\ (cross section times branching fraction) for \wpl\ ($\ell = e$ or $\mu$)
as a function of \wp\ mass.
Limits are evaluated in the context of the Sequential Standard Model (SSM), i.e.
the extended gauge model of Ref.~\cite{ssm} with the \wp\ coupling to $WZ$ set to zero.
In this model, the \wp\ has the same couplings to fermions as the SM \w~boson and
a width which increases linearly with the \wp\ mass.
A previous letter~\cite{atlas:wprime_2011-2_pub} described a similar search with a
subset (1.0~\ifb) of the data used in this study.
Here the mass range of the search is extended and the limits in the
previously-covered region are significantly improved because of the fivefold increase in integrated luminosity.
An improved lower mass limit assuming SSM coupling strength is also reported.

A search is also performed for the charged partners, denoted \wstar, of the chiral boson excitations
described in Ref.~\cite{wzstar} with theoretical motivation in
Ref.~\cite{wzstar_motivate}.
The anomalous (mag\-netic-moment type) coupling of the \wstar\ leads to kinematic distributions
significantly different from those of the \wp.
The previous search for this resonance~\cite{atlas:wprime_2010_pub} was performed using data acquired
in 2010 with an integrated luminosity less than 1\% of that used here.
The search region is expanded to both lower and higher masses and the limits  are considerably
improved in the region covered by the previous serach.
A lower mass limit is evaluated by fixing the \wstar\ coupling strengths to give the same partial decay
widths as the SSM~\wp.

The analysis presented here identifies event candidates in the electron and muon channels,
sets separate limits for \wpse\ and \wpsmu, and then combines
these assuming a common branching fraction for the two channels.
The kinematic variable used to identify the \wps\ is the transverse mass
\begin{equation}
\mt = \sqrt{ 2 \pt \met (1 - \cos \varphi_{\ell\nu})},
\end{equation}
whose distribution has a Jacobian peak and falls sharply above the resonance mass.
Here \pt\ is the lepton transverse momentum, \met\ is the magnitude of the missing transverse momentum (\mettext),
and $\varphi_{\ell\nu}$ is the angle between the \pt\ and \mettext\ vectors.
Throughout this letter, transverse refers to the plane perpendicular to the colliding beams, longitudinal
means parallel to the beams, \mytheta\ and \myphi\ are the polar and azimuthal angles with respect to the
longitudinal direction, and pseudorapidity is defined as $\eta = -\ln(\tan(\theta/2))$.

Figure~\ref{fig:wpwskin} shows the electron \eta\ and the \mt\ spectra for \wpe\ and \wse, with
$\mwp = \mws = 2.0\tev$,
from the event generation, detector simulation and reconstruction described below.
The difference in kinematic shape is evident: the \wp\ is more central in pseudorapidity
and has a sharper \mt\ spectrum.

\begin{figure*}[!ht]
  \centering
  \includegraphics[width=0.48\textwidth]{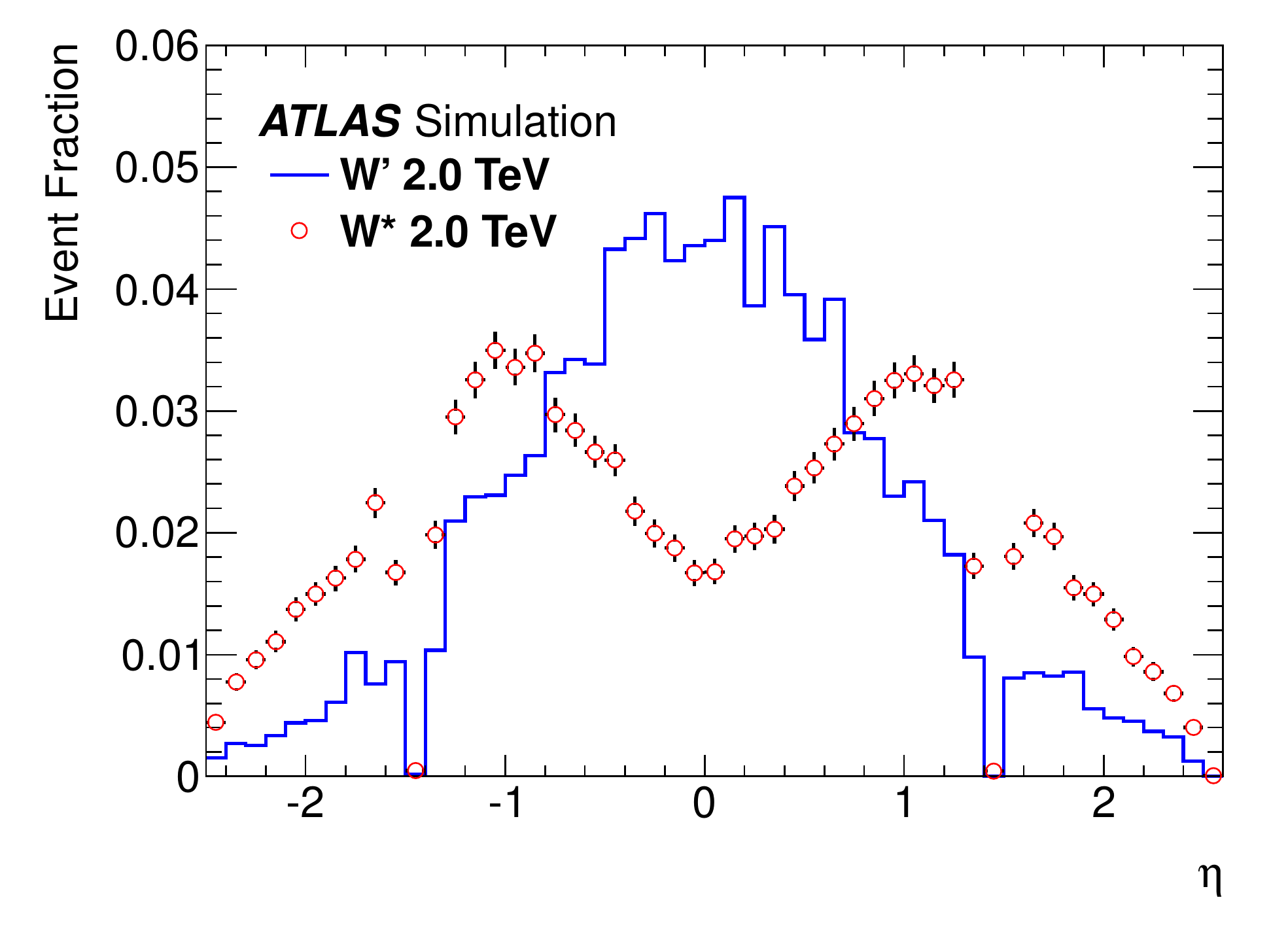}
  \includegraphics[width=0.48\textwidth]{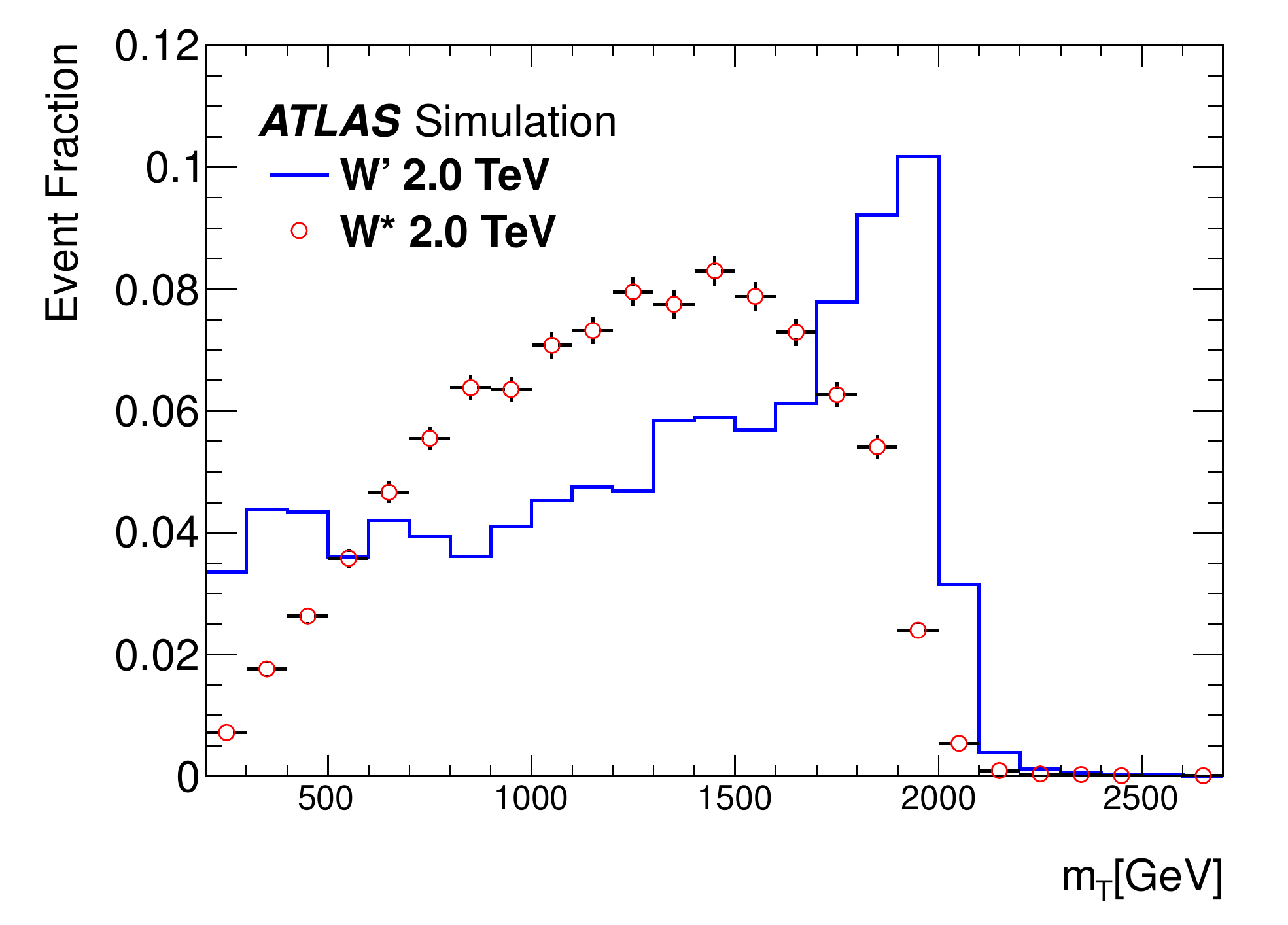}
  \caption{Reconstructed electron \eta\ (left) and \mt\ (right) distributions for
  \wpe\ and \wse\ with $\mwp = \mws = 2.0\tev$.
  All distributions are normalised to unit area.
  \label{fig:wpwskin}
  }
\end{figure*}

The main background to the \wpsl\ signal comes from the high-\mt\ tail of SM \w~boson decay to the same final state.
Other backgrounds are \z\ bosons decaying into two leptons where one lepton is not reconstructed,
\w\ or \z\ decaying to $\tau$ leptons where a $\tau$ subsequently decays to an electron or muon, and
diboson production. These are collectively referred to as the electroweak (EW) background.
In addition, there is a background contribution from \ttbar\ and single-top production which is most important for
the lowest \wp\ masses considered here, where it constitutes about 15\% of the background after event selection.
Other strong-interaction background sources, where a light or heavy hadron decays semileptonically or a jet is misidentified
as an electron, are estimated to be at most 10\% of the total background in the electron channel
and a negligible fraction in the muon channel.
These are called QCD background in the following.

\section{Detector, trigger and reconstruction}

The ATLAS detector~\cite{atlas:detector} has three major components: the inner tracking detector,
the calorimeter and the muon spectrometer.
Charged particle tracks and vertices are reconstructed with silicon pixel and silicon strip detectors
covering $|\eta|<2.5$ and straw-tube transition radiation detectors covering $|\eta| < 2.0$,
all immersed in a homogeneous 2~T magnetic field provided by a superconducting solenoid.
This tracking detector is
surrounded by a finely-segmented, hermetic calorimeter system that covers \mbox{$|\eta| < 4.9$}
and provides three-dimensional reconstruction of particle showers.
It uses liquid argon for the inner EM (electromagnetic)
compartment followed by a hadronic compartment based on scintillating tiles in the central region ($|\eta| < 1.7$)
and liquid argon for higher $|\eta|$.
Outside the calorimeter, there is a muon spectrometer with air-core toroids providing a magnetic
field, whose integral averages about 3~Tm.
The deflection of the muons in the magnetic field is measured with three layers of precision drift-tube chambers for
$|\eta| < 2.0$ and one layer of cathode-strip chambers followed by two layers of drift-tube chambers
for $2.0 < |\eta| < 2.7$.
Additional resistive-plate and thin-gap chambers provide muon triggering capability and measurement of
the $\varphi$ coordinate.

The data used in the electron channel are recorded with a trigger requiring the presence of an
EM cluster (i.e.\ an energy cluster in the EM compartment of the calorimeter)
with energy corresponding to an electron with $\pt > 80\gev$.
This substantial increase over the \pt\ threshold used in the previous analysis~\cite{atlas:wprime_2011-2_pub}
is required to maintain high efficiency (above 99\%) and keep the trigger rate at a tolerable level
for the high luminosity used to acquire the bulk of the data.
For the muon channel, matching tracks in the muon spectrometer and inner detector with combined $\pt > 22\gev$ are
used to select events. Events are also recorded if a muon with $\pt>40\gev$ is found in the muon spectrometer.
These are the same \pt\ thresholds used in the previous analysis and, despite stricter hit requirements imposed for the
higher-luminosity data, the muon trigger efficiency remains 80--90\% in the regions of interest.

Each EM cluster with \mbox{$\et > 85\gev$}
and $|\eta| < 1.37$ or $1.52 < |\eta| < 2.47$
is considered as an electron candidate if it matches an inner detector track.
The electron direction is defined as that of the reconstructed track and its energy as that of the cluster,
with a small \eta-dependent energy scale correction.
The energy resolution is 2\% for $\et \approx 50\gev$ and approaches 1\% in the high-\et\
range relevant to this analysis.
To discriminate against hadronic jets,
requirements are imposed on the lateral shower shapes in the first two layers
of the EM compartment of the calorimeter
and on the fraction of energy leaking into the hadronic compartment.
A hit in the first pixel layer is required to reduce background from photon conversions in the
inner detector material.
These requirements result in about 90\% identification efficiency for electrons with \mbox{$\et > 85 \gev$}
and a $2\times10^{-4}$ probability to falsely identify jets as electrons before isolation requirements are
imposed~\cite{atlas:wz2010}.

Muons are required to have $\pt > 25\gev$, where the
momentum of the muon is obtained by combining the inner detector and muon spectrometer measurements.
The \pt\ threshold allows the high trigger efficiency.
To ensure precise measurement of the momentum,
muons are required to have hits in all three muon layers and
are restricted to those \eta-ranges where the muon spectrometer alignment is best
understood: approximately $|\eta| < 1.0$ and $1.3 < |\eta| < 2.0$.
The average momentum resolution is about 15\% at $\pt = 1\tev$.
About 80\% of the muons in these \eta-ranges are reconstructed,
with most of the loss coming from regions with limited detector coverage.

The \mettext\ in each event is evaluated by summing over energy-calibrated physics objects
(jets, photons and leptons) and adding
corrections for calorimeter deposits away from these objects~\cite{atlas:met2011}.
This is an improvement over the previous analysis which did not include the energy calibration.

This analysis makes use of all the $\sqrt{s} = 7\tev$ data collected in 2011 
for which the relevant detector systems were operating properly.
The integrated luminosity for the data used in this study is 4.7~\ifb\ in both the electron and muon
decay channels.
The uncertainty on this measurement is 3.9\%~\cite{atlas:lumi, atlas:lumi_2011jul}.

\section{Simulation}
\label{sec:sim}

Except for the QCD background, which is measured with data, expected signal and background levels are evaluated
using simulated samples, normalised with calculated cross sections and the integrated luminosity of the data.

The \wp\ signal and the \wzbg\ boson backgrounds are generated with
\pythia~6.421~\cite{pythia} using the modified leading-order (LO) 
parton distribution functions (PDFs) of Ref.~\cite{mrst}.
\pythia\ is also used for the \wsl\ event generation, but with initial kinematics generated at 
LO with \comphep~\cite{comphep} using the CTEQ6L1 PDFs~\cite{cteq6l}.
The \ttbar\ background is generated with \mcatnlo~3.41~\cite{mcatnlo} using the CTEQ6.6~\cite{cteq} PDFs.
For all samples, final-state photon radiation is handled by \photos~\cite{photos}.
The ATLAS full detector simulation~\cite{atlas:sim} based on \geant4~\cite{geant} is used to
propagate the particles and account for the response of the detector.

The \pythia\ signal model for \wp\ has \vminusa\ SM couplings to fermions but does not include
interference between \w\ and \wp. For both \wp\ and \wstar, decays to channels other than $e\nu$ and $\mu\nu$,
including $\tau\nu$, $ud$, $sc$ and $tb$, are included in the calculation of the widths but are not explicitly
included as signal or background.
At high mass ($\mwp > 1~\tev$),
the branching fraction to each of the lepton decay channels is 8.2\%.

The \wlnu\ events are reweighted to have the NNLO
(next-to-next-to-leading-order) QCD mass dependence of
\mbox{ZWPROD~\cite{zwprod}}
following the $G_{\mu}$ scheme~\cite{horace}
and using the MSTW2008 PDFs~\cite{mstw}.
Higher-order electroweak corrections (in addition to the photon radiation included in the simulation)
are calculated using \horace~\cite{horace, horaceNC}.
In the high-mass region of interest, the electroweak corrections reduce the cross sections
by 11\% at $\mlnu = 1\tev$ and by 18\% at $\mlnu = 2\tev$.

The \wlnu\ and \zll\ cross sections are calculated at NNLO 
using FEWZ~\cite{fewz,fewz2} with the same PDFs, scheme and electroweak corrections used in
the ZWPROD event reweighting.
The \wpl\ cross sections are calculated in the same way, except
the electroweak corrections beyond final-state radiation are not included because the calculation for the SM \w\
cannot be applied directly.
The \ttbar\ cross section is calculated at approximate-NNLO~\cite{Moch:2008qy, Langenfeld:2009tc, Aliev:2010zk}
assuming a top-quark mass of 172.5~\gev.
The \wsl\ cross-section evaluation is performed with \comphep\ using the CTEQ6L1 PDFs
(i.e.\ same as the event generation).
The signal and most important background cross sections are listed in Table~\ref{tab:xsec}.

Cross-section uncertainties for \wpl\ and the \wzbg~\cite{atlas:wz2010} and \ttbar~\cite{atlas:ttbar2010}
backgrounds are estimated from the MSTW2008 PDF error sets,
the difference between the MSTW2008 and CTEQ6.6 PDFs, and variation of renormalization and factorization scales
by a factor of two.
The estimates from the three sources are combined in quadrature.
Most of the net uncertainty comes from the PDF error sets and the MSTW-CTEQ difference, in roughly equal proportion.
The \wsl\ cross-section uncertainties are evaluated with the CTEQ61~\cite{cteq6l} PDF error sets.

\begin{table}[!tb]
\caption{
Calculated values of \xbr\ for \wpl, \wsl\ and the leading backgrounds.
The value for $\ttbar\rightarrow\ell X$ includes all final states with at least one
lepton ($e$, $\mu$ or $\tau$).
The others are exclusive and are used for both $\ell=e$ and $\ell=\mu$.
All calculations are NNLO except \wstar\ which is LO and
\ttbar\ which is approximate-NNLO.
}
\label{tab:xsec}
\begin{center}
\begin{tabular}{l|c|l}
\hline
\hline
        & Mass        &             \\
Process &       [\gev] & $\xbr$ [pb] \\
\hline
\multirow{14}{*}{\wpl} &
   \phantom{0}300 & \phantom{00}130.5      \\
 & \phantom{0}400 & \phantom{000}41.6      \\
 & \phantom{0}500 & \phantom{000}17.25     \\
 & \phantom{0}600 & \phantom{0000}8.27     \\
 & \phantom{0}750 & \phantom{0000}3.20     \\
 & \phantom{}1000 & \phantom{0000}0.837    \\
 & \phantom{}1250 & \phantom{0000}0.261    \\
 & \phantom{}1500 & \phantom{0000}0.0887   \\
 & \phantom{}1750 & \phantom{0000}0.0325   \\
 & \phantom{}2000 & \phantom{0000}0.0126   \\
 & \phantom{}2250 & \phantom{0000}0.00526  \\
 & \phantom{}2500 & \phantom{0000}0.00235  \\
 & \phantom{}2750 & \phantom{0000}0.001156 \\
 & \phantom{}3000 & \phantom{0000}0.000643 \\
\hline
\multirow{12}{*}{\wsl} &
   \phantom{0}400 & \phantom{000}29.6      \\
 & \phantom{0}500 & \phantom{000}12.6      \\
 & \phantom{0}750 & \phantom{0000}2.34     \\
 & \phantom{}1000 & \phantom{0000}0.610    \\
 & \phantom{}1250 & \phantom{0000}0.188    \\
 & \phantom{}1500 & \phantom{0000}0.0636   \\
 & \phantom{}1750 & \phantom{0000}0.0226   \\
 & \phantom{}2000 & \phantom{0000}0.00819  \\
 & \phantom{}2250 & \phantom{0000}0.00299  \\
 & \phantom{}2500 & \phantom{0000}0.000109 \\
 & \phantom{}2750 & \phantom{0000}0.000391 \\
 & \phantom{}3000 & \phantom{0000}0.000138 \\
\hline
\wlnu\ & & \phantom{}10460 \\
\zgll\               & & \multirow{2}{*}{\phantom{00}989} \\
($m_{\zg}>60 \gev$)  & & \\
 \ttbarl &  & \phantom{000}89.4 \\
\hline
\hline
\end{tabular}
\end{center}
\end{table}

\section{Event selection}

The primary vertex for each event is required to have at least three tracks
with $\pt > 0.4$~\gev\ and to have a longitudinal distance less than 200~mm
from the centre of the collision region.
Due to the high luminosity, there are an average of more than ten additional
interactions per event in the data used for this analysis.
The primary vertex is defined to be the one with the highest summed track $\pt^2$.
Spurious tails in \mettext, arising from calorimeter noise and other detector problems are suppressed by checking the quality
of each reconstructed jet and discarding events where any jet has a shape indicating such problems,
following Ref.~\cite{atlas:plhc:jet_cleaning}.
In addition, the inner detector track associated with the electron or muon is required to be compatible
with originating from the primary vertex, specifically to have transverse distance of closest approach
$|d_0|<1\mm$ and longitudinal distance at this point $|z_0|<5\mm$ in the electron channel.
For the muon channel, the requirements are $|d_0|<0.2\mm$ and $|z_0|<1\mm$.
Events are required to have exactly one candidate electron or one candidate muon satisfying
these requirements.

To suppress the QCD background, the lepton is required to be isolated. In the electron channel,
the isolation energy is measured with the calorimeter in a cone $\Delta R < 0.4$
($\Delta R \equiv \sqrt{(\Delta \eta)^2 + (\Delta \varphi)^2}$) around the electron track,
and the requirement is \mbox{$\sumet < 9 \gev$,}
where the sum includes all calorimeter energy clusters in the cone excluding the core energy deposited by the electron.
The sum is  
corrected to account for additional interactions and leakage of the electron energy outside this core.
In the muon channel, the isolation energy is measured using inner detector tracks with $\pttrack > 1$~\gev\ in a cone
$\Delta R < 0.3$ around the muon track.
The isolation requirement is
\mbox{$\sumpttrack < 0.05~\pt$},
where the muon track is excluded from the sum.
The scaling of the threshold with the muon \pt\ 
reduces efficiency losses due to radiation from the muon at high \pt.

\mettextcap\ thresholds are imposed to further suppress the background
from QCD and $W+$jets (events where the SM $W$ recoils against hadronic jets).
In both channels, the threshold used for the charged lepton
\pt\ is also applied to the \mettext:
$\met > 85 \gev$ for the electron channel and $\met > 25 \gev$ for the muon channel.

The above constitute the event preselection requirements.
An \mt\ threshold varying with \wp\ or \wstar\ mass and decay channel is applied after preselection to
establish the final event counts.

In the electron channel, the QCD background is estimated from data using the \emph{ABCD}
technique~\cite{atlas:photon2010} with the isolation energy and \mettext\ serving as discriminants.
Consistent results are obtained using the \emph{inverted isolation} technique described
in Ref.~\cite{atlas:wprime_2010_pub}.

The QCD background for the muon channel is evaluated using the
\emph{matrix method}~\cite{atlas:ttbar2010}.
This background is less than 1\% of the total
background, and so it is neglected in the following.

The same reconstruction and event selection are applied to both data and simulated samples.
Figure~\ref{fig:final_mt} shows the charged lepton \pt, missing \et, and \mt\ spectra
for events with $\mt > 200\gev$ in each channel after event preselection.
The data, the expected background, and three examples of \wp\ signals at different masses are shown.
The \mt\ threshold, which is below that used in all of the final selections,
discriminates against the $W+$jets and QCD backgrounds.
The \mt\ spectra for the data and
expected background are consistent within statistical and systematic uncertainties.

Table~\ref{tab:nbgone} shows the contributions to the background
for $\mt>794 \gev$, the region used to search for a \wp\ with a mass of 1000~\gev.
The \wlnu\ background dominates and the background for the electron channel is higher than that
for muons because of the difference in acceptance.

\begin{figure*}[!ht]
  \centering
  \includegraphics[width=0.48\textwidth]{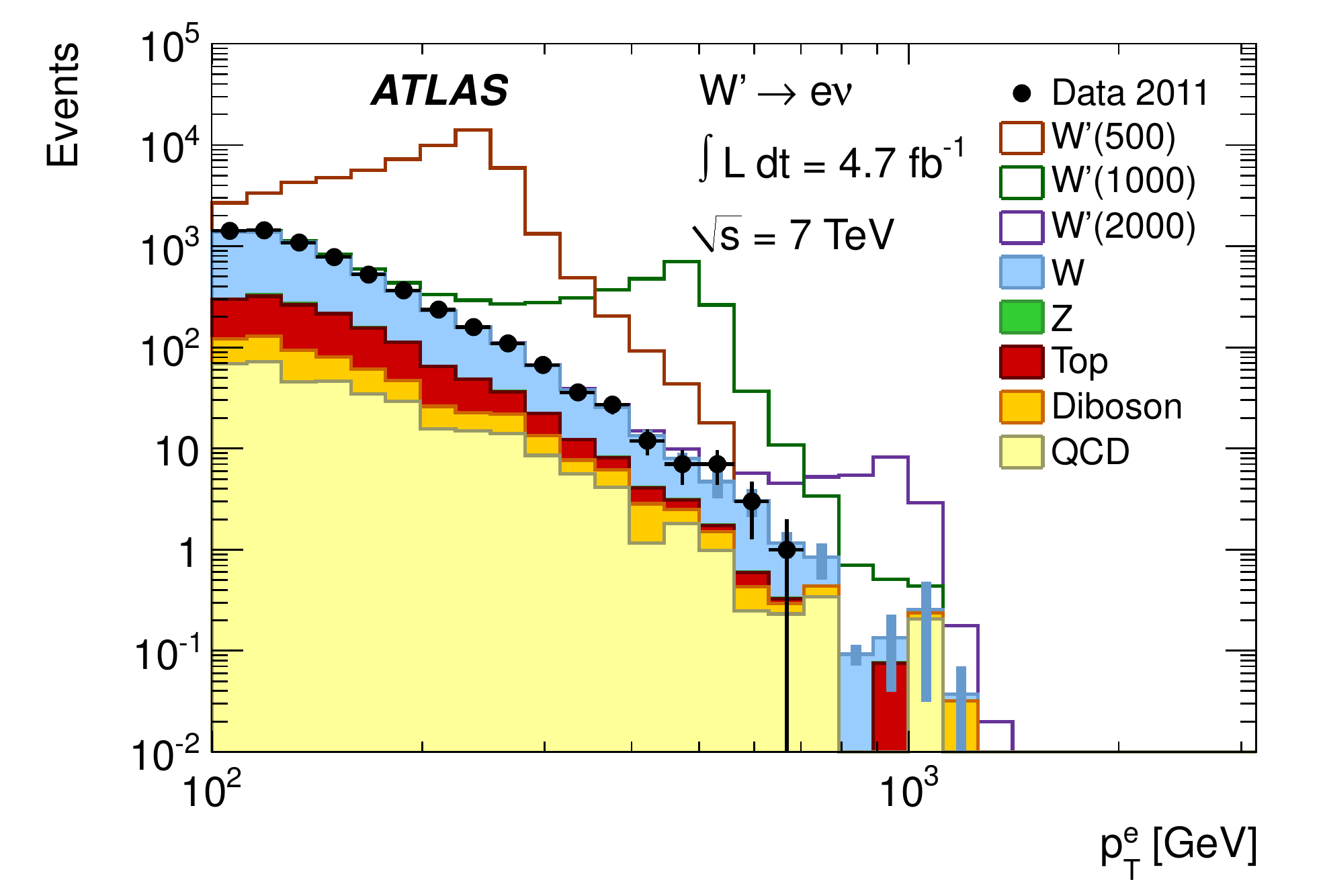}
  \includegraphics[width=0.48\textwidth]{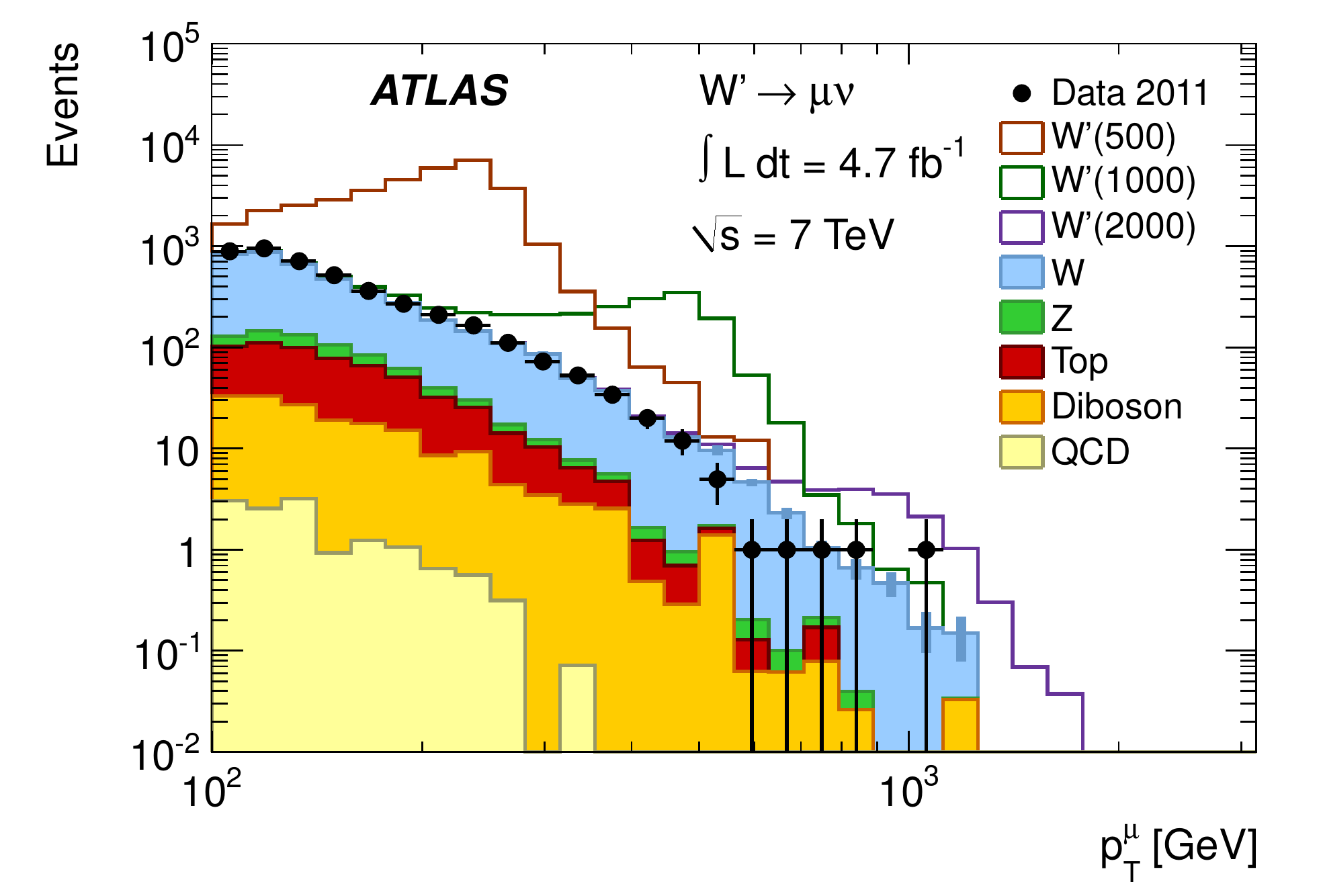}
  \includegraphics[width=0.48\textwidth]{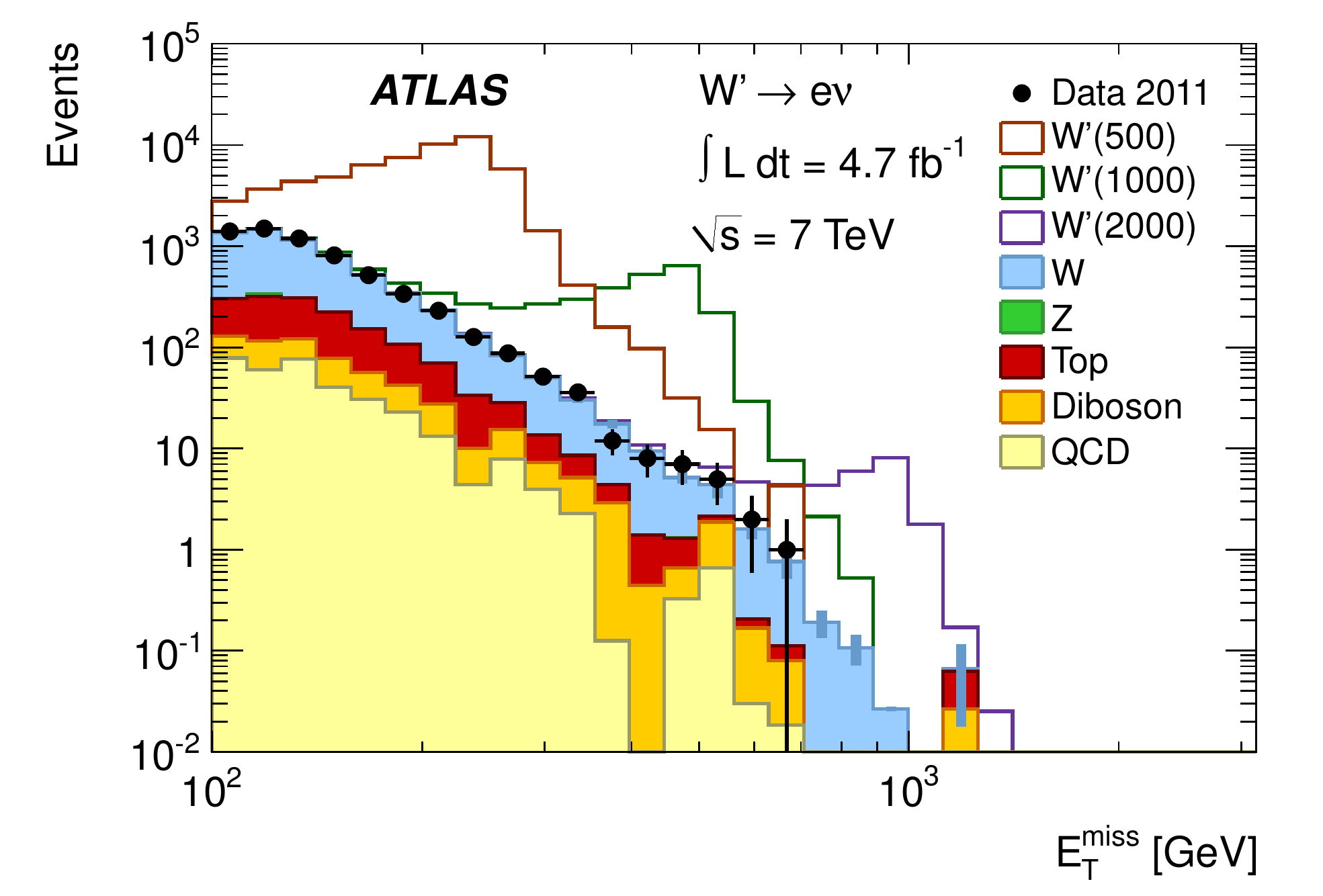}
  \includegraphics[width=0.48\textwidth]{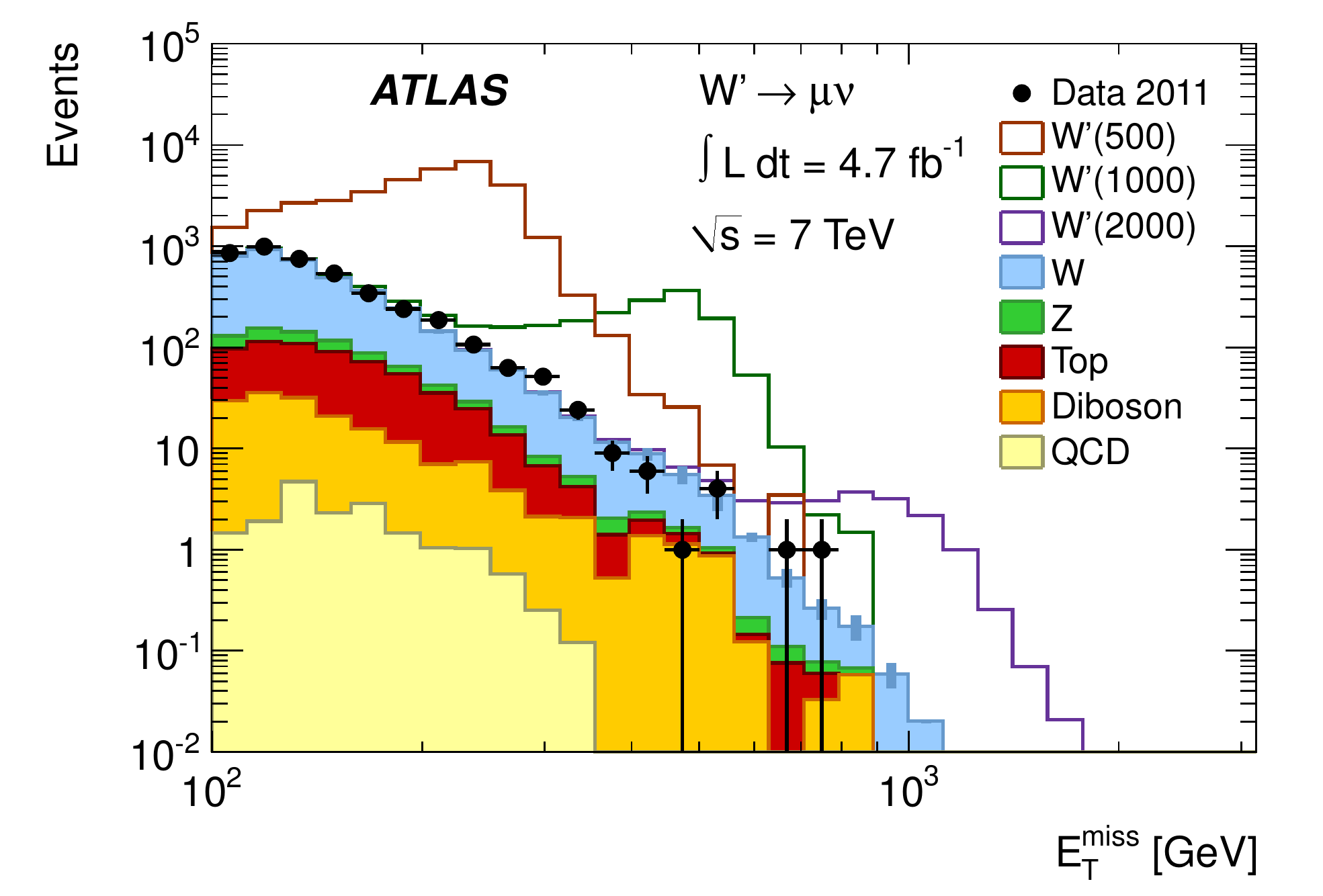}
  \includegraphics[width=0.48\textwidth]{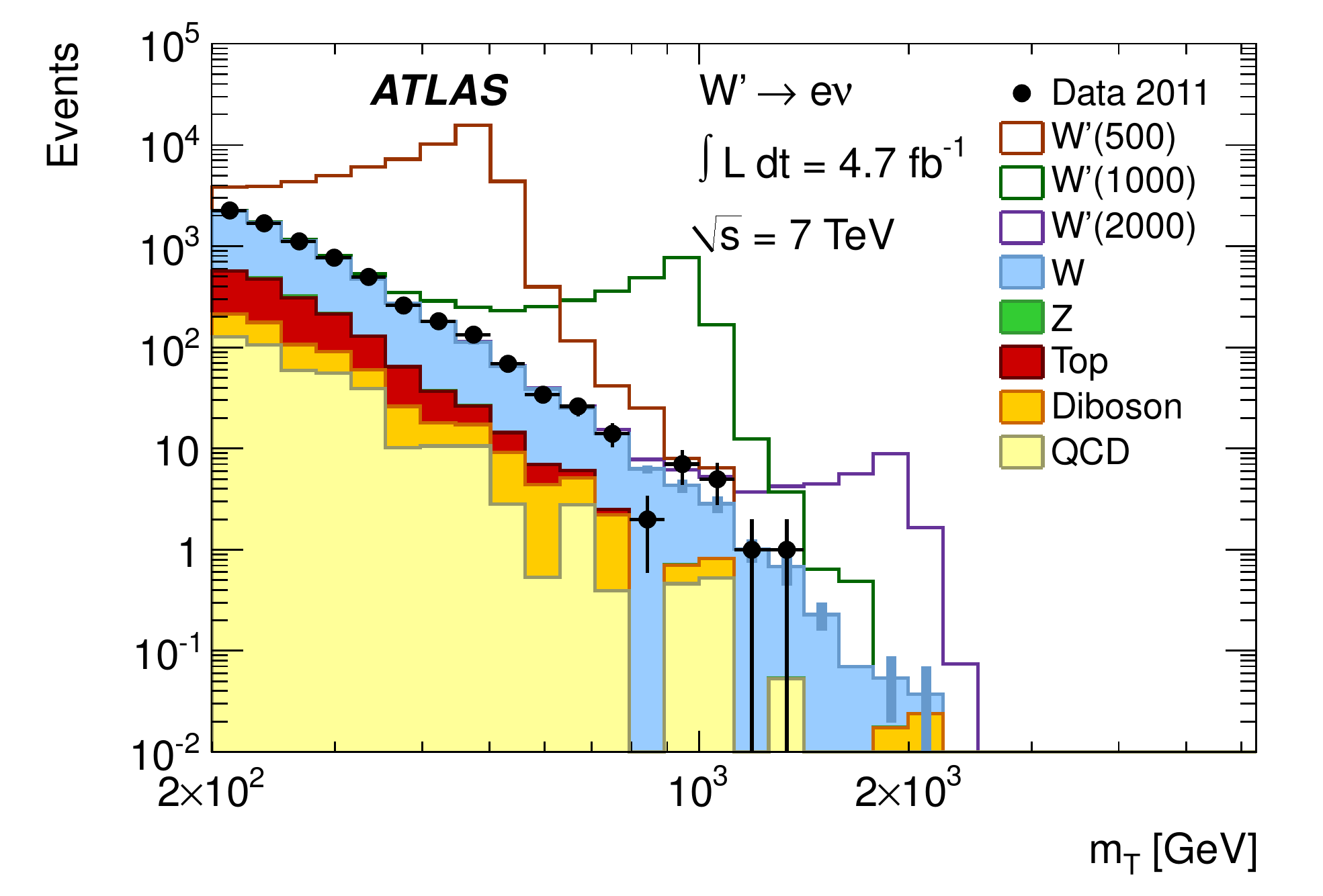}
  \includegraphics[width=0.48\textwidth]{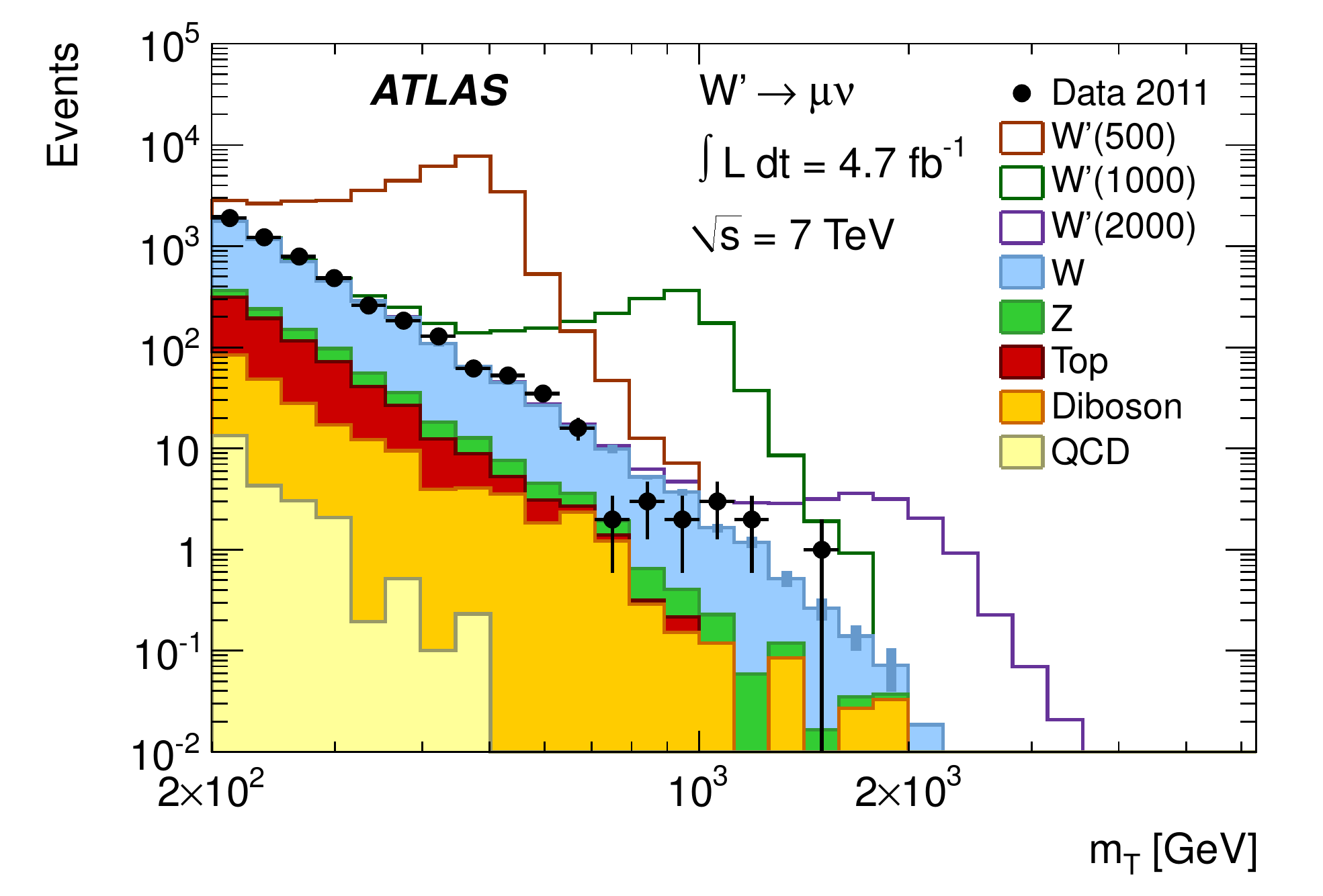}
  \caption{Spectra of charged lepton \pt\ (top), \mettext\ (centre) and \mt\ (bottom)
  for the electron (left) and muon (right) channels for events with $\mt > 200\gev$ after event preselection.
  The points represent data and the filled histograms show the stacked backgrounds.
  Open histograms are \wpl\ signals added to the background with masses in \gev\ indicated in parentheses in the legend. 
  The QCD backgrounds estimated from data are also shown.
  The signal and other background samples are normalised using the integrated luminosity of the data
  and the NNLO (approximate-NNLO for \ttbar) cross sections listed in Table~\ref{tab:xsec}.
  The error bars on the data and background sums are statistical, i.e the latter do not include the systematic
  uncertainties used in the statistical analysis.
  \label{fig:final_mt}
  }
\end{figure*}

\begin{table}[!bt]
\caption{
Expected numbers of events from the various background sources in each decay channel
for \mbox{$\mt>794\gev$}, the region used to
search for a \wp\ with a mass of 1000~\gev\ in the electron and muon channels.
The \wlnu\ and \zll\ entries include the expected contributions from the $\tau$-lepton.
The uncertainties are those from the Monte Carlo statistics.
}
\label{tab:nbgone}
\begin{center}
\begin{tabular}{l | l|l}
\hline
\hline
                & \multicolumn{1}{c|}{$e\nu$} & \multicolumn{1}{c}{$\mu\nu$} \\
\hline
\tspace
\wlnu           & \phantom{}14.2\phantom{00} $\pm$ 0.5         & \phantom{}11.2\phantom{0}  $\pm$ 0.5  \\
\zll            & \phantom{0}0.022\phantom{} $\pm$ 0.001       & \phantom{0}0.76\phantom{}  $\pm$ 0.01 \\
diboson         & \phantom{0}1.2\phantom{00} $\pm$ 0.2         & \phantom{0}0.71\phantom{} $\pm$ 0.15  \\
\ttbar          & \phantom{0}0.24\phantom{0} $\pm$ 0.11        & \phantom{0}0.09\phantom{} $\pm$ 0.05  \\
QCD             & \phantom{0}0.8\phantom{00} $\pm$ 0.3         & \phantom{00000}$-$                    \\
\hline
\tspace
Total           & \phantom{}16.5\phantom{00} $\pm$ 0.6         & \phantom{}12.8\phantom{0} $\pm$ 0.5 \\
\hline
\hline
\end{tabular}
\end{center}
\end{table}

\section{Statistical analysis and systematics}

Discovery significance and \xbr\ limits are evaluated
independently for \wp\ and \wstar\
following the same procedure as
for the previous analysis~\cite{atlas:wprime_2011-2_pub}.
The observed number of events \nobs\ is the count after final selection including the
requirement $\mt > \mtmin$, with that threshold chosen separately
for each mass and decay channel to maximize sensitivity.
A Bayesian posterior probability distribution for the signal \xbr\ is evaluated
with a Poisson likelihood at each mass for each decay channel and
for the combination of the two channels.
A positive, flat prior is used for the signal \xbr, and Gaussian distributions
are used for the three nuisance parameters: \effsig, the efficiency to select signal events,
\nbg, the expected number of background events and \lint, the integrated luminosity.
For each observed posterior, an ensemble of expected posteriors is generated assuming
no signal and the same prior distributions for \nbg\ and \lint.

Each of the observed posteriors is used to evaluate an observed limit on \xbr, and the
ensemble of expected posteriors provides the corresponding expected limit distribution.
All limits are at 95\% CL (credibility level).
Discovery significance is assessed from the fraction of the expected
posteriors that are more signal-like than the observation.

The values and uncertainties for \effsig\ are presented in Tables~\ref{tab:wpeff} and~\ref{tab:wseff},
and those for \nbg\ and \nobs\ in Table~\ref{tab:nbg}.
The \effsig\ tables also give the predicted numbers of signal events, \nsig, with their uncertainties accounting
for the uncertainties in both \effsig\ and the cross-section calculations.

\begin{table}[htbp]
\caption{Event selection efficiencies for the \wpe\ and \wpmu\ searches.
The first three columns are the \wp\ mass, \mt\ threshold and decay channel.
The next two are the signal selection efficiency, \effsig, and the prediction
for the number of signal events, \nsig, obtained with this efficiency.
The uncertainty on \nsig\ includes contributions from the uncertainty on the
cross sections but not from that on the integrated luminosity.
\label{tab:wpeff}
}
\begin{center}
\begin{tabular}{rrr| r@{ $\pm$ }r r@{ $\pm$ }r}
\hline
\hline
 \mwp   & \mtmin & \\
{[}\gev] & [\gev] &     & \multicolumn{2}{c}{\effsig} & \multicolumn{2}{c}{\nsig} \\
\hline
\hline
\multirow{2}{*}{300} & \multirow{2}{*}{251}
        & $e\nu$   & 0.288 & 0.023 & 176000\phantom{.00} & 19000\phantom{.00} \\
 &      & $\mu\nu$ & 0.186 & 0.016 & 114000\phantom{.00} & 13000\phantom{.00} \\
\hline
\multirow{2}{*}{400} & \multirow{2}{*}{355}
        & $e\nu$   & 0.237 & 0.023 &  46200\phantom{.00} &  5600\phantom{.00} \\
 &      & $\mu\nu$ & 0.153 & 0.018 &  30000\phantom{.00} &  4100\phantom{.00} \\
\hline
\multirow{2}{*}{500} & \multirow{2}{*}{447}
        & $e\nu$   & 0.237 & 0.023 &  19200\phantom{.00} &  2300\phantom{.00} \\
 &      & $\mu\nu$ & 0.145 & 0.019 &  11700\phantom{.00} &  1800\phantom{.00} \\
\hline
\multirow{2}{*}{600} & \multirow{2}{*}{501}
        & $e\nu$   & 0.307 & 0.024 &  11900\phantom{.00} &  1300\phantom{.00} \\
 &      & $\mu\nu$ & 0.195 & 0.017 &   7600\phantom{.00} &   900\phantom{.00} \\
\hline
\multirow{2}{*}{750} & \multirow{2}{*}{631}
        &   $e\nu$ & 0.297 & 0.023 &   4470\phantom{.00} &   470\phantom{.00} \\
 &      & $\mu\nu$ & 0.189 & 0.016 &   2840\phantom{.00} &   320\phantom{.00} \\
\hline
\multirow{2}{*}{1000}  & \multirow{2}{*}{794}
        &   $e\nu$ & 0.339 & 0.023 &   1330\phantom{.00} &   130\phantom{.00} \\
 &      & $\mu\nu$ & 0.223 & 0.015 &    877\phantom{.00} &    90\phantom{.00} \\
\hline
\multirow{2}{*}{1250} & \multirow{2}{*}{1000}
        &   $e\nu$ & 0.323 & 0.024 &    395\phantom{.00} &    47\phantom{.00} \\
 &      & $\mu\nu$ & 0.212 & 0.019 &    259\phantom{.00} &    34\phantom{.00} \\
\hline
\multirow{2}{*}{1500} & \multirow{2}{*}{1122}
        &   $e\nu$ & 0.351 & 0.026 &    146\phantom{.00} &    20\phantom{.00} \\
 &      & $\mu\nu$ & 0.237 & 0.021 &     99\phantom{.00} &    14\phantom{.00} \\
\hline
\multirow{2}{*}{1750} & \multirow{2}{*}{1413}
        &   $e\nu$ & 0.280 & 0.024 &     42.7\phantom{0} &     6.8\phantom{0} \\
 &      & $\mu\nu$ & 0.179 & 0.024 &     27.3\phantom{0} &     5.2\phantom{0} \\
\hline
\multirow{2}{*}{2000} & \multirow{2}{*}{1413}
        &   $e\nu$ & 0.317 & 0.025 &     18.8\phantom{0} &     3.2\phantom{0} \\
 &      & $\mu\nu$ & 0.215 & 0.022 &     12.7\phantom{0} &     2.3\phantom{0} \\
\hline
\multirow{2}{*}{2250} & \multirow{2}{*}{1413}
        &   $e\nu$ & 0.315 & 0.022 &      7.8\phantom{0} &     1.5\phantom{0} \\
 &      & $\mu\nu$ & 0.218 & 0.017 &      5.4\phantom{0} &     1.0\phantom{0} \\
\hline
\multirow{2}{*}{2500} & \multirow{2}{*}{1413}
        &   $e\nu$ & 0.276 & 0.024 &      3.1\phantom{0} &     1.4\phantom{0} \\
 &      & $\mu\nu$ & 0.184 & 0.024 &      2.0\phantom{0} &     1.0\phantom{0} \\
\hline
\multirow{2}{*}{2750} & \multirow{2}{*}{1413}
        &   $e\nu$ & 0.217 & 0.020 &      1.18\phantom{} &     0.59\phantom{} \\
 &      & $\mu\nu$ & 0.149 & 0.020 &      0.81\phantom{} &     0.41\phantom{} \\
\hline
\multirow{2}{*}{3000} & \multirow{2}{*}{1413}
        &   $e\nu$ & 0.143 & 0.027 &      0.43\phantom{} &     0.25\phantom{} \\
 &      & $\mu\nu$ & 0.106 & 0.031 &      0.32\phantom{} &     0.20\phantom{} \\
\hline
\hline
\end{tabular}
\end{center}
\end{table}

\begin{table}[htbp]
\caption{Event selection efficiencies for the \wse\ and \wsmu\ searches.
The first three columns are the \wstar\ mass, \mt\ threshold and decay channel.
The next two are the signal selection efficiency, \effsig, and the prediction
for the number of signal events, \nsig, obtained with this efficiency.
The uncertainty on \nsig\ includes contributions from the uncertainty on the
cross sections but not from that on the integrated luminosity.
\label{tab:wseff}
}
\begin{center}
\begin{tabular}{rrr| r@{ $\pm$ }r r@{ $\pm$ }r}
\hline
\hline
 \mws   & \mtmin & \\
{[}\gev] & [\gev] &     & \multicolumn{2}{c}{\effsig} & \multicolumn{2}{c}{\nsig} \\
\hline
\hline
\multirow{2}{*}{400} & \multirow{2}{*}{316}
        & $e\nu$   & 0.189 & 0.021 &  26300\phantom{.000} &  3200\phantom{.000} \\
 &      & $\mu\nu$ & 0.118 & 0.020 &  16400\phantom{.000} &  2900\phantom{.000} \\
\hline
\multirow{2}{*}{500} & \multirow{2}{*}{398}
        & $e\nu$   & 0.182 & 0.020 &  10800\phantom{.000} &  1300\phantom{.000} \\
 &      & $\mu\nu$ & 0.114 & 0.021 &   6740\phantom{.000} &  1300\phantom{.000} \\
\hline
\multirow{2}{*}{750} & \multirow{2}{*}{562}
        &   $e\nu$ & 0.224 & 0.021 &   2460\phantom{.000} &   270\phantom{.000} \\
 &      & $\mu\nu$ & 0.143 & 0.019 &   1570\phantom{.000} &   230\phantom{.000} \\
\hline
\multirow{2}{*}{1000} & \multirow{2}{*}{708}
        &   $e\nu$ & 0.267 & 0.022 &    766\phantom{.000} &    83\phantom{.000} \\
 &      & $\mu\nu$ & 0.172 & 0.017 &    493\phantom{.000} &    60\phantom{.000} \\
\hline
\multirow{2}{*}{1250}
 &  891 &   $e\nu$ & 0.254 & 0.021 &    225\phantom{.000} &    26\phantom{.000} \\
 &  794 & $\mu\nu$ & 0.216 & 0.015 &    192\phantom{.000} &    21\phantom{.000} \\
\hline
\multirow{2}{*}{1500}
 & 1122 &   $e\nu$ & 0.212 & 0.021 &     63.5\phantom{00} &     9.0\phantom{00} \\
 & 1000 & $\mu\nu$ & 0.192 & 0.016 &     57.5\phantom{00} &     7.5\phantom{00} \\
\hline
\multirow{2}{*}{1750} & \multirow{2}{*}{1122}
        &   $e\nu$ & 0.330 & 0.023 &     35.0\phantom{00} &     5.0\phantom{00} \\
 &      & $\mu\nu$ & 0.208 & 0.016 &     22.1\phantom{00} &     3.2\phantom{00} \\
\hline
\multirow{2}{*}{2000} & \multirow{2}{*}{1413}
        &   $e\nu$ & 0.258 & 0.021 &      9.9\phantom{00} &     1.7\phantom{00} \\
 &      & $\mu\nu$ & 0.156 & 0.018 &      6.0\phantom{00} &     1.2\phantom{00} \\
\hline
\multirow{2}{*}{2250} & \multirow{2}{*}{1413}
        &   $e\nu$ & 0.338 & 0.024 &      4.8\phantom{00} &     1.0\phantom{00} \\
 &      & $\mu\nu$ & 0.211 & 0.016 &      2.97\phantom{0} &     0.63\phantom{0} \\
\hline
\multirow{2}{*}{2500} & \multirow{2}{*}{1413}
        &   $e\nu$ & 0.397 & 0.025 &      2.03\phantom{0} &     0.53\phantom{0} \\
 &      & $\mu\nu$ & 0.241 & 0.016 &      1.23\phantom{0} &     0.32\phantom{0} \\
\hline
\multirow{2}{*}{2750} & \multirow{2}{*}{1413}
        &   $e\nu$ & 0.449 & 0.027 &      0.83\phantom{0} &     0.28\phantom{0} \\
 &      & $\mu\nu$ & 0.260 & 0.016 &      0.48\phantom{0} &     0.16\phantom{0} \\
\hline
\multirow{2}{*}{3000} & \multirow{2}{*}{1413}
        &   $e\nu$ & 0.475 & 0.029 &      0.31\phantom{0} &     0.13\phantom{0} \\
 &      & $\mu\nu$ & 0.276 & 0.016 &      0.179\phantom{} &     0.077\phantom{} \\
\hline
\hline
\end{tabular}
\end{center}
\end{table}

\begin{table}[htbp]
\caption{Background levels and observed counts for the \wpl\ and \wsl\ searches
in both the electron and muon channels.
The first two columns are the \mt\ threshold and decay channel, followed by
the expected number of background
events, \nbg, and the number of events observed in data, \nobs.
The uncertainty on \nbg\ includes contributions from the uncertainties on the
cross sections but not from that on the integrated luminosity.
\label{tab:nbg}
}
\begin{center}
\begin{tabular}{rr| r@{ $\pm$ }r r}
\hline
\hline
 \mtmin &  & \multicolumn{2}{c}{} \\
{[}\gev] &            & \multicolumn{2}{c}{\nbg}       & \nobs \\
\hline
\hline
 \multirow{2}{*}{ 251} &   $e\nu$ & 3190\phantom{.00} & 260\phantom{.00} & 3105 \\
                       & $\mu\nu$ & 1950\phantom{.00} & 190\phantom{.00} & 2023 \\
\hline
 \multirow{2}{*}{ 316} &   $e\nu$ & 1240\phantom{.00} & 100\phantom{.00} & 1229 \\
                       & $\mu\nu$ &  773\phantom{.00} &  72\phantom{.00} &  750 \\
\hline
 \multirow{2}{*}{ 355} &   $e\nu$ &  761\phantom{.00} &  64\phantom{.00} &  734 \\
                       & $\mu\nu$ &  492\phantom{.00} &  44\phantom{.00} &  491 \\
\hline
 \multirow{2}{*}{ 398} &   $e\nu$ &  467\phantom{.00} &  39\phantom{.00} &  474 \\
                       & $\mu\nu$ &  285\phantom{.00} &  26\phantom{.00} &  307 \\
\hline
 \multirow{2}{*}{ 447} &   $e\nu$ &  277\phantom{.00} &  24\phantom{.00} &  293 \\
                       & $\mu\nu$ &  178\phantom{.00} &  15\phantom{.00} &  179 \\
\hline
 \multirow{2}{*}{ 501} &   $e\nu$ &  164\phantom{.00} &  14\phantom{.00} &  159 \\
                       & $\mu\nu$ &  113\phantom{.00} &  10\phantom{.00} &  117 \\
\hline
 \multirow{2}{*}{ 562} &   $e\nu$ &  95.8\phantom{.0} &   8.4\phantom{0} &   90 \\
                       & $\mu\nu$ &  66.2\phantom{.0} &   5.8\phantom{0} &   64 \\
\hline
 \multirow{2}{*}{ 631} &   $e\nu$ &   54.5\phantom{0} &   5.2\phantom{0} &   56 \\
                       & $\mu\nu$ &   40.0\phantom{0} &   3.7\phantom{0} &   29 \\
\hline
 \multirow{2}{*}{ 708} &   $e\nu$ &   30.7\phantom{0} &   3.0\phantom{0} &   30 \\
                       & $\mu\nu$ &   22.7\phantom{0} &   2.2\phantom{0} &   13 \\
\hline
 \multirow{2}{*}{ 794} &   $e\nu$ &   16.5\phantom{0} &   1.7\phantom{0} &   16 \\
                       & $\mu\nu$ &   12.8\phantom{0} &   1.4\phantom{0} &   11 \\
\hline
                  891  &  $\e\nu$ &    9.0\phantom{0} &   1.0\phantom{0} &   14 \\
\hline
 \multirow{2}{*}{1000} &   $e\nu$ &    5.15\phantom{} &   0.69\phantom{} &    7 \\
                       & $\mu\nu$ &    3.86\phantom{} &   0.58\phantom{} &    6 \\
\hline
 \multirow{2}{*}{1122} &   $e\nu$ &    2.57\phantom{} &   0.42\phantom{} &    2 \\
                       & $\mu\nu$ &    2.21\phantom{} &   0.34\phantom{} &    3 \\
\hline
 \multirow{2}{*}{1413} &   $e\nu$ &    0.64\phantom{} &   0.18\phantom{} &    0 \\
                       & $\mu\nu$ &    0.51\phantom{} &   0.12\phantom{} &    1 \\
\hline
\hline
\end{tabular}
\end{center}
\end{table}

The maximum value for the \wpl\ signal selection efficiency is at $\mwp = 1500\gev$.
For lower masses, the efficiency falls because the relative \mt\ threshold, $\mtmin/\mwp$,
is increased to reduce the background level.
For higher masses, the efficiency falls because a large fraction of the cross section goes via
off-shell production with $\mlnu \muchless \mwp$.
This effect is not seen for \wsl\ because its derivative couplings~\cite{wzstar}
suppress off-shell production at low mass.

The fraction of fully simulated signal events that pass the event selection and are above the
\mt\ threshold provides the initial estimate of \effsig\ for each channel and mass.
For \wp, small corrections are then made to account for the difference in acceptance at NNLO
(obtained from FEWZ) and that in the LO simulation.
These vary from a 10\% increase for $\mwp = 500\gev$ to an 11\% decrease for $\mwp = 2500\gev$.
Contributions from \mbox{\wptau} with the \taulep\ decaying leptonically have been neglected.
These would increase the \wp\ signal strength by 3--4\% for the highest masses.
The background level is estimated for each mass by summing the EW and \ttbar\ event counts
from simulation, and adding the small QCD contribution in the electron channel.

The uncertainties on \effsig, \nbg\ and \lint\ account for experimental and theoretical systematic
effects as well as the statistics of the simulation samples.
The uncertainty on \lint\ is included separately to allow for the correlation between signal and background.
The experimental systematic uncertainties include efficiencies for the electron or muon trigger, reconstruction
and selection.
Lepton momentum and \mettext\ response, characterised by scale and resolution, are also included.
Most of these performance metrics are measured at relatively low \pt\ and their values are extrapolated to
the high-\pt\ regime relevant to this analysis.
The uncertainties in these extrapolations are included but their contributions are small compared to the
total uncertainty on \effsig\ or \nbg.
The uncertainty on the QCD background estimate also contributes to the background-level uncertainties for the electron channel.
Theoretical uncertainties include those from the cross-section calculations
(see Section~\ref{sec:sim}) and from the \wp\ acceptance corrections.
The values for the uncertainties are similar to those obtained in the previous analysis.
Table~\ref{tab:syst_summary} summarizes the uncertainties on the event selection efficiencies 
and background levels for the \mbox{\wpl} signal with $\mwp=1500\gev$ using $\mt>1122\gev$.

\begin{table}[!ht]
\caption{
  Relative uncertainties on the event selection efficiency and background level for
  a \wp\ with a mass of 1500~\gev.
  The efficiency uncertainties include contributions from the trigger, reconstruction and event selection.
  The cross-section uncertainty for \effsig\ is that assigned to the acceptance correction described in the text.
  The cross-section uncertainty on \nbg\ is that from the cross-section calculations.
  The last row gives the total uncertainties.
\label{tab:syst_summary}
}
\begin{center}
\begin{tabular}{l|rr|rr}
\hline
\hline
 & \multicolumn{2}{c|}{\effsig} & \multicolumn{2}{c}{\nbg} \\
 Source                      &  \multicolumn{1}{c}{$e\nu$}  & \multicolumn{1}{c|}{$\mu\nu$}
                             &  \multicolumn{1}{c}{$e\nu$}  & \multicolumn{1}{c}{$\mu\nu$} \\
\hline
 Efficiency                  &                  5\%  &                   2\%  &   4\% &                  2\%  \\
 Energy/momentum resolution  & \multicolumn{1}{c}{-} &                   1\%  &   3\% & \multicolumn{1}{c}{-} \\
 Energy/momentum scale       &                  2\%  & \multicolumn{1}{c|}{-} &   4\% & \multicolumn{1}{c}{-} \\
 \mettextcap                 & \multicolumn{1}{c}{-} & \multicolumn{1}{c|}{-} &   2\% &                  4\%  \\
 QCD background              & \multicolumn{1}{c}{-} & \multicolumn{1}{c|}{-} &   4\% & \multicolumn{1}{c}{-} \\
 Monte Carlo statistics      &                  5\%  &                   9\%  &  10\% &                  9\%  \\
 Cross section (shape/level) &                  3\%  &                   3\%  &  12\% &                 12\%  \\
\hline
 Total                       &                  7\% &                    9\%  &  17\% &                 16\% \\
\hline
\hline
\end{tabular}
\end{center}
\end{table}

\section{Results}

None of the observations for any mass point in either channel or their combination shows an
excess with significance above three sigma,
so there is no evidence for the observation of \wpl\ or \wsl.
Tables~\ref{tab:limits_xbr} and~\ref{tab:limits_xbr_ws} and Fig.~\ref{fig:limits_xbr} present the
95\% CL observed limits on \xbr\ for both \wpl\ and \wsl\
in the electron channel, the muon channel and their combination.
The tables also give the limits obtained without systematic uncertainties and with various subsets.
The uncertainties on the signal efficiency have very little effect on the final limits, and the
background-level and luminosity uncertainties are important only for the lowest masses.
The figure also shows the expected limits and the theoretical \xbr\ for an SSM \wp\
and for a \wstar\ with quark and gluon coupling strengths normalised to
reproduce the \wp\ width.

The intersection between the central theoretical prediction and the observed limits
provides the 95\% CL lower limits on the mass.
Table~\ref{tab:limits_mass} presents the expected and observed \wp\  and \wstar\ mass limits for
the electron and muon decay channels and their combination.

\begin{figure*}[!htbp]
  \centering
  \includegraphics[width=0.49\textwidth]{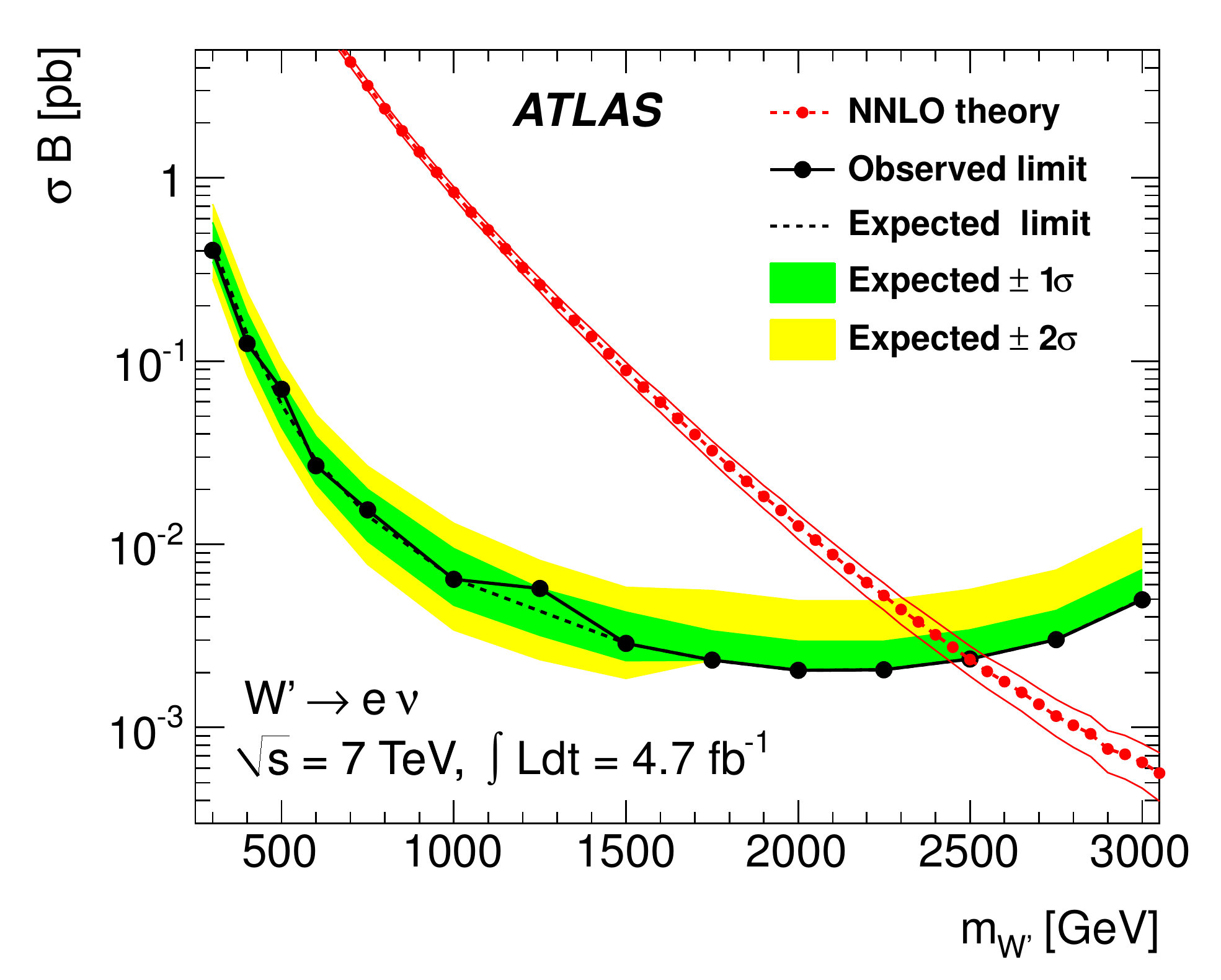}
  \includegraphics[width=0.49\textwidth]{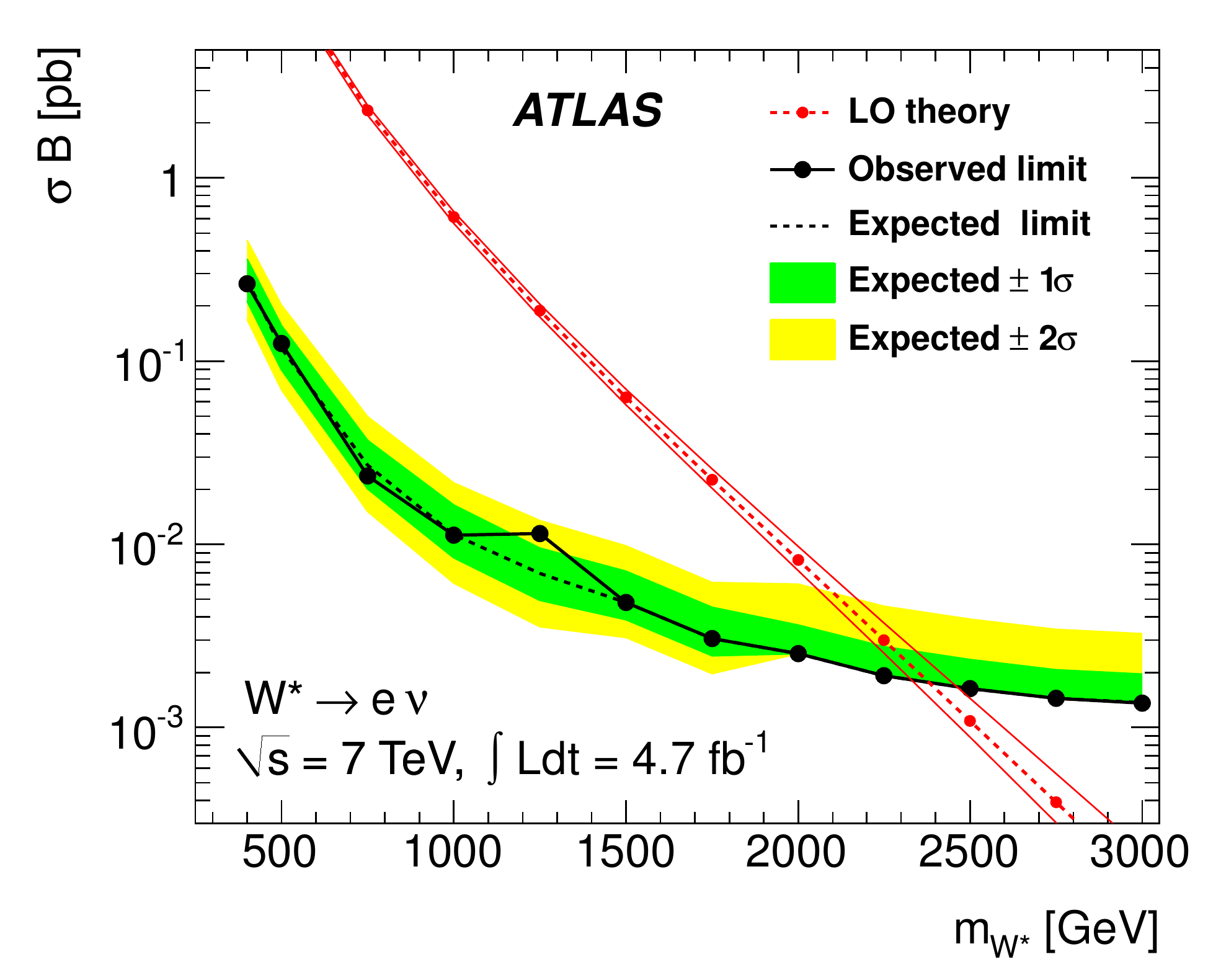}
  \includegraphics[width=0.49\textwidth]{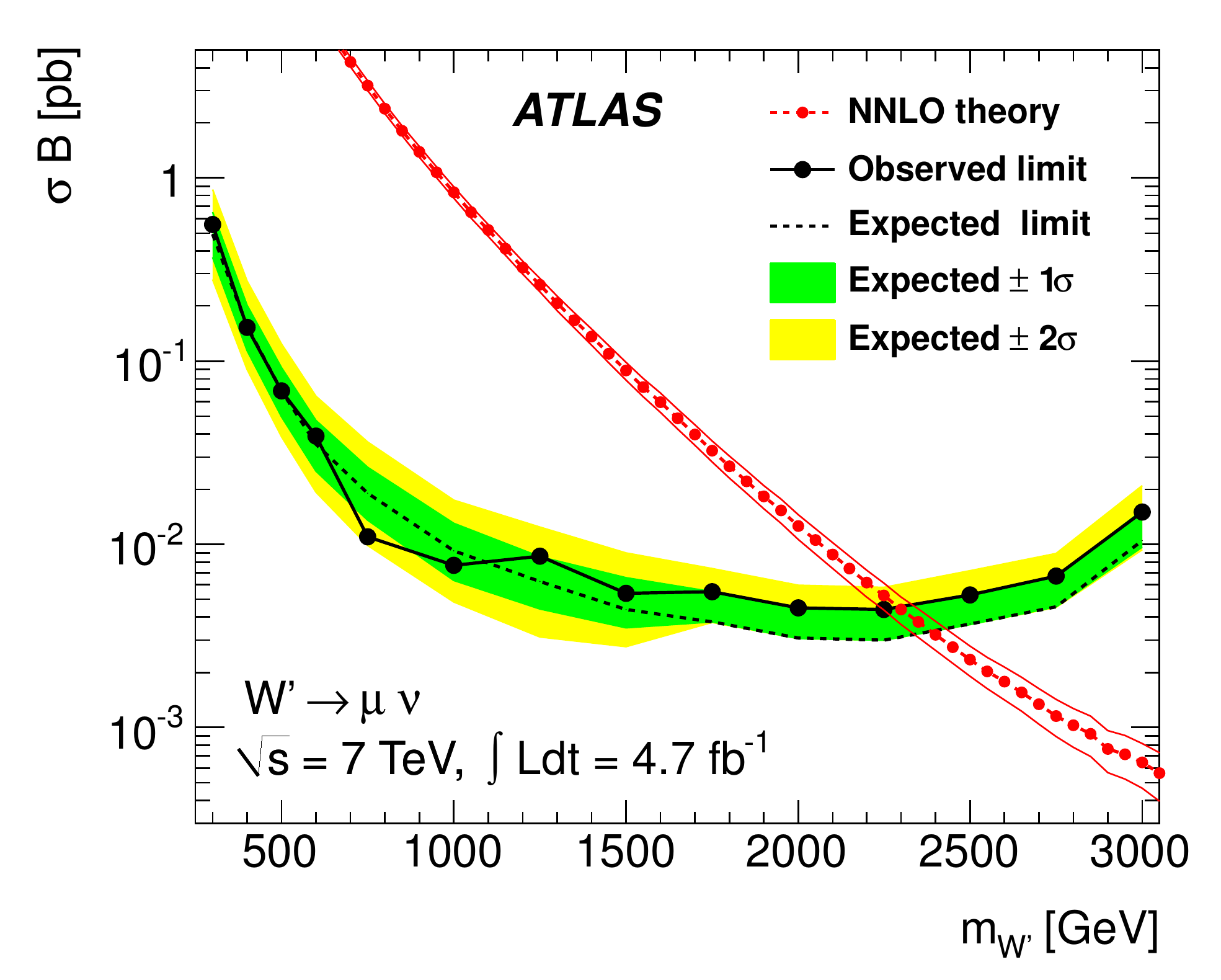}
  \includegraphics[width=0.49\textwidth]{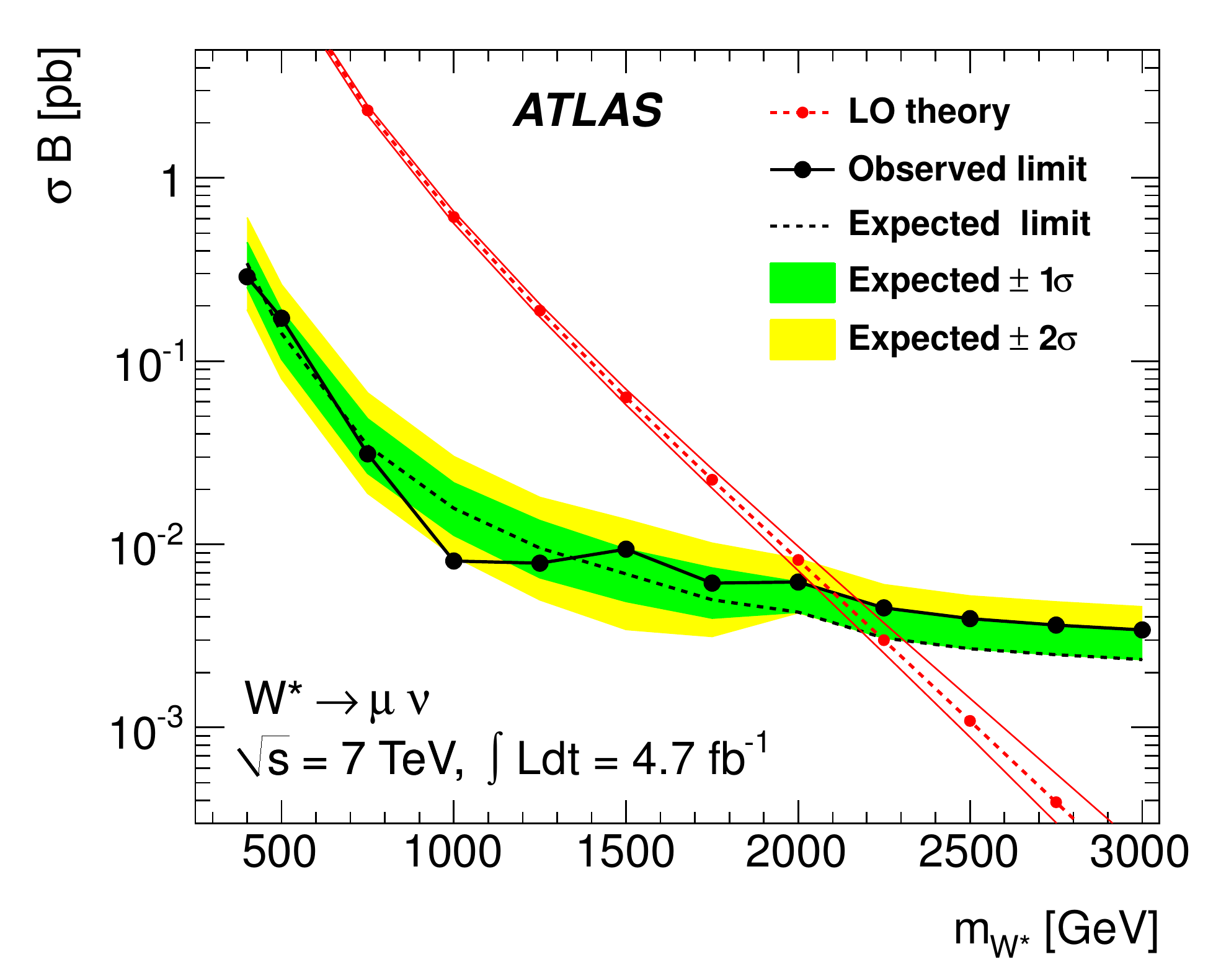}
  \includegraphics[width=0.49\textwidth]{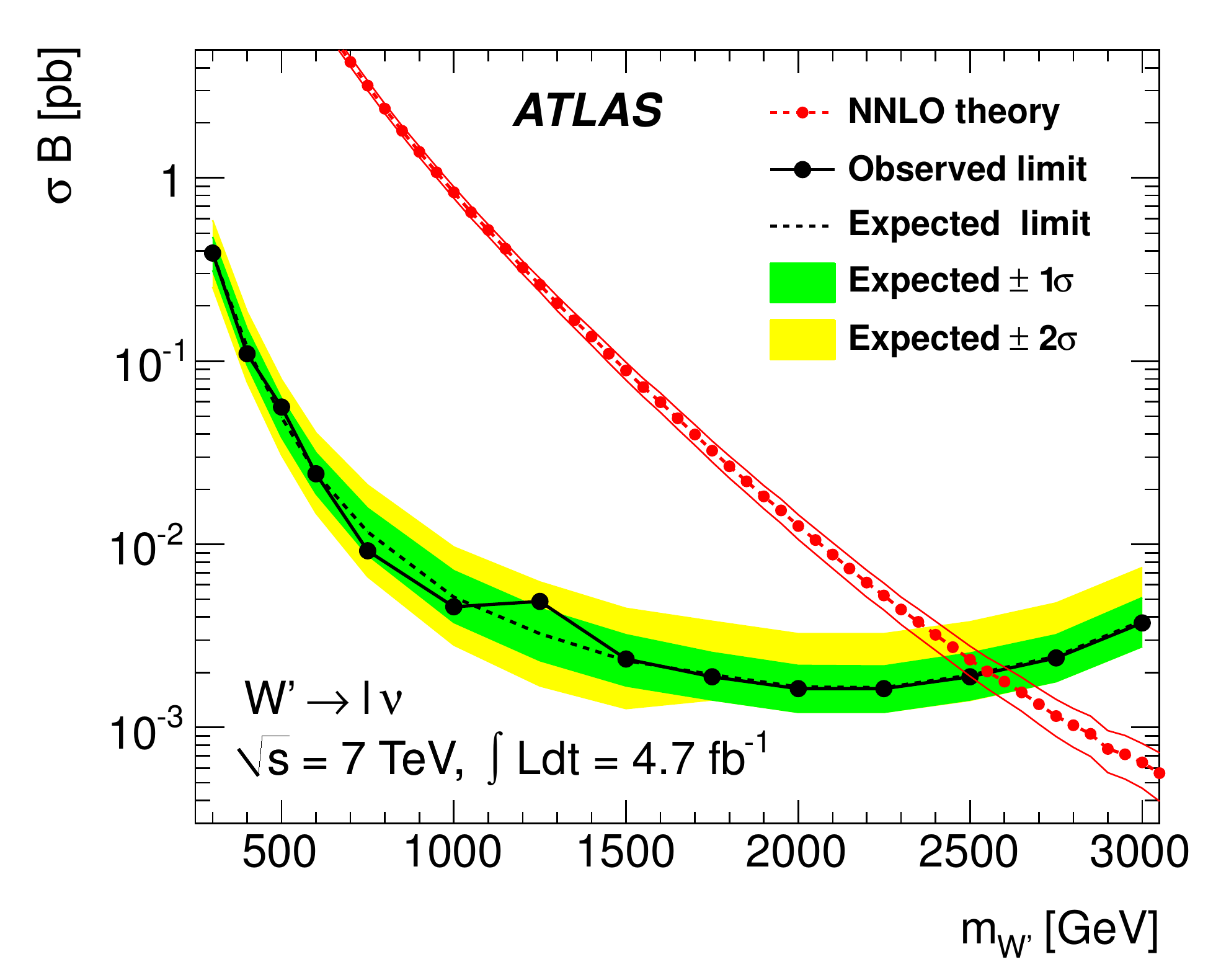}
  \includegraphics[width=0.49\textwidth]{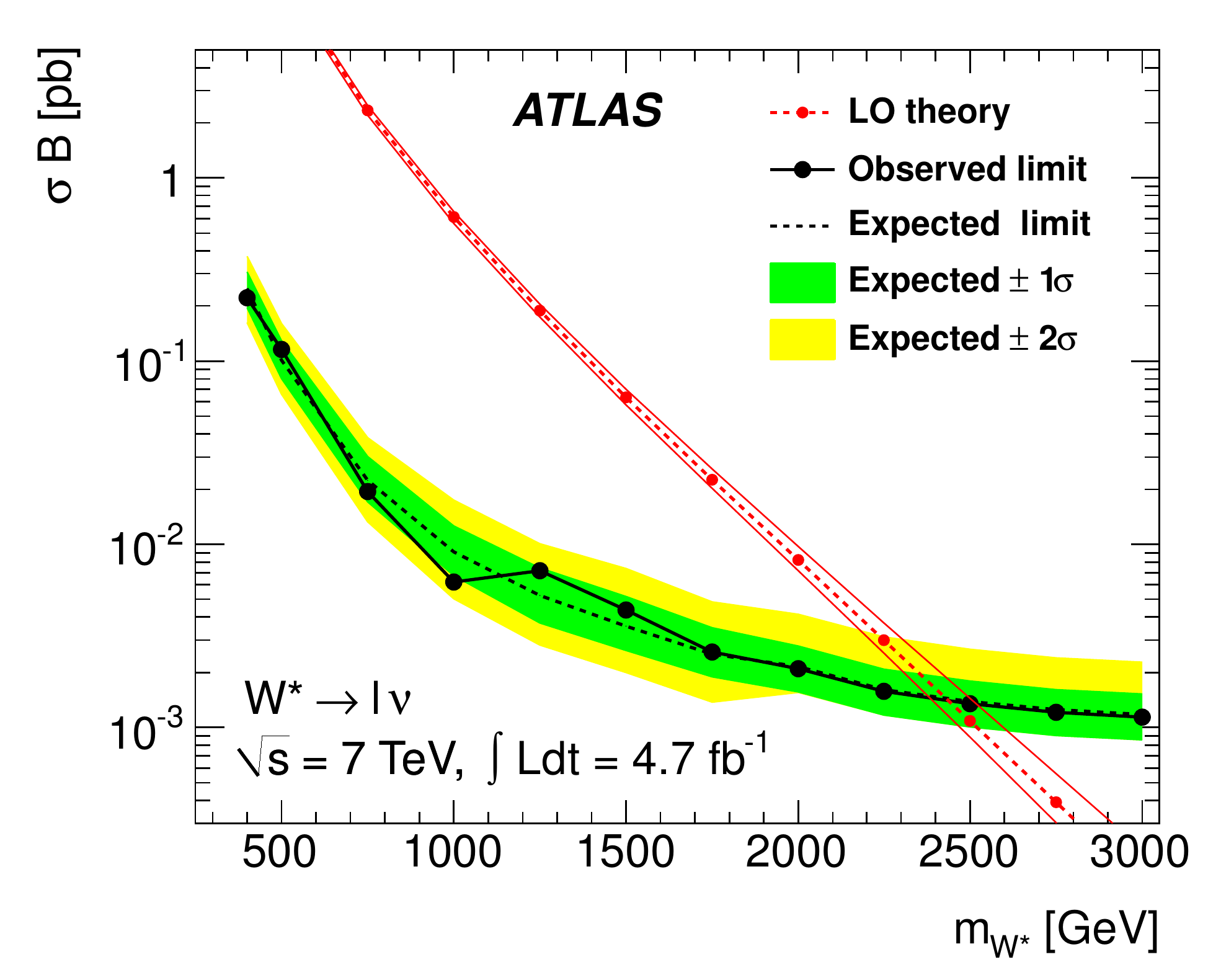}
  \caption{
  Expected and observed limits on \xbr\ for \wpl\ (left) and \wsl\ (right) in
  the electron channel (top), muon channel (centre) and combined (bottom)
  assuming the same branching fraction for both channels.
  The calculated values for \xbr\ (NNLO for \wp\ and LO for \wstar) and their uncertainties are also shown.
  \label{fig:limits_xbr}
  }
\end{figure*}

\begin{table}[!p]
\caption{Observed upper limits on \xbr\
         for \wpe, \wpmu\ and the combination of the two.
         The first two columns are the \wp\ mass and decay channel.
         The following columns are
         the 95\% CL limits with headers indicating the nuisance parameters
         for which uncertainties are included: S for the event selection efficiency (\effsig),
         B for the background level (\nbg), and L for the integrated luminosity (\lint).
         These values neglect correlations between the two channels for the combined limit.
         The only important correlation, that from background cross section, is included
         in the column $\mathrm{SB_cL}$.
         The last column in each row (SBL for $e$ and $\mu$ and $\mathrm{SB_cL}$ for $e\mu$)
         is the final limit (including all systematic uncertainties) for the mass listed in
         the first column.
         These are the limits shown in Fig.~\ref{fig:limits_xbr} (left).
\label{tab:limits_xbr}
}

\begin{center}
\begin{tabular}{rr|rrrrr}
\hline
\hline
 \mwp &      & \multicolumn{4}{c}{95\% CL limit on \xbr\ [fb]} \\
 {[}\gev] &            & none & S & SB  & SBL & $\mathrm{SB_cL}$ \\
\hline
\multirow{3}{*}{300}
&       $e$ &    50\phantom{.0} &  51\phantom{.0} & 356\phantom{.0} & 500\phantom{.0} & \\
&       $\mu$ & 173\phantom{.0} & 179\phantom{.0} & 514\phantom{.0} & 557\phantom{.0} &  \\
&       $e\mu$ & 61\phantom{.0} &  62\phantom{.0} & 295\phantom{.0} & 329\phantom{.0} & 389\phantom{.0} \\
\hline
\multirow{3}{*}{400}
&       $e$ &    36\phantom{.0} &  37\phantom{.0} & 111\phantom{.0} & 124\phantom{.0} & \\
&       $\mu$ &  62\phantom{.0} &  65\phantom{.0} & 140\phantom{.0} & 153\phantom{.0} &  \\
&       $e\mu$ & 30\phantom{.0} &  30\phantom{.0} &  84\phantom{.0} &  92\phantom{.0} & 110\phantom{.0} \\
\hline
\multirow{3}{*}{500}
&       $e$ &    43\phantom{.0} &  44\phantom{.0} &  65\phantom{.0} &  70\phantom{.0} &  \\
&       $\mu$ &  42\phantom{.0} &  44\phantom{.0} &  64\phantom{.0} &  69\phantom{.0} &  \\
&       $e\mu$ & 32\phantom{.0} &  32\phantom{.0} &  47\phantom{.0} &  50\phantom{.0} &  56\phantom{.0} \\
\hline
\multirow{3}{*}{600}
&       $e$ &    16\phantom{.0} &  17\phantom{.0} &  25\phantom{.0} &  27\phantom{.0} &  \\
&       $\mu$ &  28\phantom{.0} &  29\phantom{.0} &  36\phantom{.0} &  39\phantom{.0} &  \\
&       $e\mu$ & 14\phantom{.0} &  14\phantom{.0} &  21\phantom{.0} &  22\phantom{.0} &  24\phantom{.0} \\
\hline
\multirow{3}{*}{750}
&       $e$ &    12\phantom{.0} &  13\phantom{.0} &  15\phantom{.0} &  15\phantom{.0} &  \\
&       $\mu$ &   9.0\phantom{} &   9.2\phantom{} &  11\phantom{.0} &  11\phantom{.0} &  \\
&       $e\mu$ &  6.8\phantom{} &   6.8\phantom{} &   8.1\phantom{} &   8.4\phantom{} &   9.2\phantom{} \\
\hline
\multirow{3}{*}{1000}
&       $e$ &     5.6\phantom{} &   6.0\phantom{} &   6.3\phantom{} &   6.5\phantom{} &  \\
&       $\mu$ &   7.1\phantom{} &   7.2\phantom{} &   7.5\phantom{} &   7.7\phantom{} &  \\
&       $e\mu$ &  4.1\phantom{} &   4.1\phantom{} &   4.4\phantom{} &   4.4\phantom{} &   4.6\phantom{} \\
\hline
\multirow{3}{*}{1250}
&       $e$ &     5.5\phantom{} &   5.5\phantom{} &   5.6\phantom{} &   5.7\phantom{} &  \\
&       $\mu$ &   8.2\phantom{} &   8.4\phantom{} &   8.5\phantom{} &   8.6\phantom{} &  \\
&       $e\mu$ &  4.7\phantom{} &   4.7\phantom{} &   4.8\phantom{} &   4.9\phantom{} &   4.9\phantom{} \\
\hline
\multirow{3}{*}{1500}
&       $e$ &     2.8\phantom{} &   2.8\phantom{} &   2.9\phantom{} &   2.9\phantom{} &  \\
&       $\mu$ &   5.2\phantom{} &   5.4\phantom{} &   5.4\phantom{} &   5.4\phantom{} &  \\
&       $e\mu$ &  2.3\phantom{} &   2.3\phantom{} &   2.3\phantom{} &   2.4\phantom{} &   2.4\phantom{} \\
\hline
\multirow{3}{*}{1750}
&       $e$ &     2.3\phantom{} &   2.3\phantom{} &   2.3\phantom{} &   2.3\phantom{} &  \\
&       $\mu$ &   5.2\phantom{} &   5.5\phantom{} &   5.5\phantom{} &   5.5\phantom{} &  \\
&       $e\mu$ &  1.9\phantom{} &   1.9\phantom{} &   1.9\phantom{} &   1.9\phantom{} &   1.9\phantom{} \\
\hline
\multirow{3}{*}{2000}
&       $e$ &     2.0\phantom{} &   2.0\phantom{} &   2.0\phantom{} &   2.1\phantom{} &  \\
&       $\mu$ &   4.3\phantom{} &   4.4\phantom{} &   4.5\phantom{} &   4.5\phantom{} &  \\
&       $e\mu$ &  1.6\phantom{} &   1.6\phantom{} &   1.6\phantom{} &   1.6\phantom{} &   1.6\phantom{} \\
\hline
\multirow{3}{*}{2250}
&       $e$ &     2.0\phantom{} &   2.1\phantom{} &   2.1\phantom{} &   2.1\phantom{} &  \\
&       $\mu$ &   4.2\phantom{} &   4.3\phantom{} &   4.3\phantom{} &   4.4\phantom{} &  \\ 
&       $e\mu$ &  1.6\phantom{} &   1.6\phantom{} &   1.6\phantom{} &   1.6\phantom{} &   1.6\phantom{} \\
\hline
\multirow{3}{*}{2500}
&       $e$ &     2.3\phantom{} &   2.4\phantom{} &   2.4\phantom{} &   2.4\phantom{} &  \\
&       $\mu$ &   5.0\phantom{} &   5.3\phantom{} &   5.3\phantom{} &   5.3\phantom{} &  \\ 
&       $e\mu$ &  1.9\phantom{} &   1.9\phantom{} &   1.9\phantom{} &   1.9\phantom{} &   1.9\phantom{} \\
\hline
\multirow{3}{*}{2750}
&       $e$ &     2.9\phantom{} &   3.0\phantom{} &   3.0\phantom{} &   3.0\phantom{} &  \\
&       $\mu$ &   6.2\phantom{} &   6.6\phantom{} &   6.6\phantom{} &   6.7\phantom{} &  \\ 
&       $e\mu$ &  2.3\phantom{} &   2.4\phantom{} &   2.4\phantom{} &   2.4\phantom{} &   2.4\phantom{} \\
\hline
\multirow{3}{*}{3000}
&       $e$ &     4.5\phantom{} &   5.0\phantom{} &   5.0\phantom{} &   5.0\phantom{} &  \\
&       $\mu$ &   8.7\phantom{} &  15\phantom{.0} &   15\phantom{0} &  15\phantom{.0} &  \\
&       $e\mu$ &  3.5\phantom{} &   3.7\phantom{} &   3.7\phantom{} &   3.7\phantom{} &   3.7\phantom{} \\
\hline
\hline
\end{tabular}
\end{center}
\end{table}

\begin{table}[!p]
\caption{Observed upper limits on \xbr\
         for \wse, \wsmu\ and the combination of the two.
         The columns are as for Table~\ref{tab:limits_xbr}.
         The final (rightmost) limits are shown in Fig.~\ref{fig:limits_xbr} (right).
\label{tab:limits_xbr_ws}
}

\begin{center}
\begin{tabular}{rr|rrrrrrr}
\hline
\hline
 \mws &      & \multicolumn{4}{c}{95\% CL limit on \xbr\ [fb]} \\
 ~[\gev] &            & none & S & SB  & SBL & $\mathrm{SB_cL}$ \\
\hline
\multirow{3}{*}{400}
&       $e$ &     68\phantom{.0} &  71\phantom{.0} & 236\phantom{.0} & 264\phantom{.0} & \\
&       $\mu$ &   68\phantom{.0} &  75\phantom{.0} & 263\phantom{.0} & 289\phantom{.0} & \\
&       $e\mu$ &  47\phantom{.0} &  48\phantom{.0} & 167\phantom{.0} & 186\phantom{.0} & 222\phantom{.0} \\
\hline
\multirow{3}{*}{500}
&       $e$ &     57\phantom{.0} &  60\phantom{.0} & 114\phantom{.0} & 125\phantom{.0} & \\
&       $\mu$ &   93\phantom{.0} & 106\phantom{.0} & 160\phantom{.0} & 171\phantom{.0} & \\
&       $e\mu$ &  57\phantom{.0} &  58\phantom{.0} &  96\phantom{.0} & 104\phantom{.0} & 116\phantom{.0} \\
\hline
\multirow{3}{*}{750}
&       $e$ &     16\phantom{.0} &  17\phantom{.0} &  22\phantom{.0} &  24\phantom{.0} & \\
&       $\mu$ &   23\phantom{.0} &  25\phantom{.0} &  30\phantom{.0} &  31\phantom{.0} & \\
&       $e\mu$ &  13\phantom{.0} &  13\phantom{.0} &  17\phantom{.0} &  18\phantom{.0} &  19\phantom{.0} \\
\hline
\multirow{3}{*}{1000}
&       $e$ &     10\phantom{.0} &  10\phantom{.0} &  11\phantom{.0} &  11\phantom{.0} & \\
&       $\mu$ &    7.0\phantom{} &   7.2\phantom{} &   7.8\phantom{} &   8.1\phantom{} & \\
&       $e\mu$ &   5.0\phantom{} &   5.1\phantom{} &   5.6\phantom{} &   5.8\phantom{} &   6.2\phantom{} \\
\hline
\multirow{3}{*}{1250}
&       $e$ &     11\phantom{.0} &  11\phantom{.0} &  11\phantom{.0} &  11\phantom{.0} & \\
&       $\mu$ &    7.3\phantom{} &   7.4\phantom{} &   7.8\phantom{} &   7.9\phantom{} & \\
&       $e\mu$ &   6.7\phantom{} &   6.7\phantom{} &   6.9\phantom{} &   7.0\phantom{} &   7.2\phantom{} \\
\hline
\multirow{3}{*}{1500}
&       $e$ &      4.6\phantom{} &   4.7\phantom{} &   4.8\phantom{} &   4.8\phantom{} & \\
&       $\mu$ &    9.0\phantom{} &   9.2\phantom{} &   9.3\phantom{} &   9.4\phantom{} & \\
&       $e\mu$ &   4.2\phantom{} &   4.3\phantom{} &   4.3\phantom{} &   4.3\phantom{} &   4.4\phantom{} \\
\hline
\multirow{3}{*}{1750}
&       $e$ &      3.0\phantom{} &   3.0\phantom{} &   3.0\phantom{} &   3.0\phantom{} & \\
&       $\mu$ &    6.0\phantom{} &   6.1\phantom{} &   6.1\phantom{} &   6.2\phantom{} & \\
&       $e\mu$ &   2.5\phantom{} &   2.5\phantom{} &   2.6\phantom{} &   2.6\phantom{} &   2.6\phantom{} \\
\hline
\multirow{3}{*}{2000}
&       $e$ &      2.5\phantom{} &   2.5\phantom{} &   2.5\phantom{} &   2.5\phantom{} & \\
&       $\mu$ &    5.9\phantom{} &   6.2\phantom{} &   6.2\phantom{} &   6.2\phantom{} & \\
&       $e\mu$ &   2.1\phantom{} &   2.1\phantom{} &   2.1\phantom{} &   2.1\phantom{} &   2.1\phantom{} \\
\hline
\multirow{3}{*}{2250}
&       $e$ &      1.9\phantom{} &   1.9\phantom{} &   1.9\phantom{} &   1.9\phantom{} & \\
&       $\mu$ &    4.4\phantom{} &   4.5\phantom{} &   4.5\phantom{} &   4.5\phantom{} & \\
&       $e\mu$ &   1.6\phantom{} &   1.6\phantom{} &   1.6\phantom{} &   1.6\phantom{} &   1.6\phantom{} \\
\hline
\multirow{3}{*}{2500}
&       $e$ &      1.5\phantom{} &   1.5\phantom{} &   1.5\phantom{} &   1.5\phantom{} & \\
&       $\mu$ &    3.8\phantom{} &   3.9\phantom{} &   3.9\phantom{} &   3.9\phantom{} & \\
&       $e\mu$ &   1.3\phantom{} &   1.3\phantom{} &   1.3\phantom{} &   1.4\phantom{} &   1.4\phantom{} \\
\hline
\multirow{3}{*}{2750}
&       $e$ &      1.4\phantom{} &   1.4\phantom{} &   1.4\phantom{} &   1.4\phantom{} & \\
&       $\mu$ &    3.6\phantom{} &   3.6\phantom{} &   3.6\phantom{} &   3.6\phantom{} & \\
&       $e\mu$ &   1.2\phantom{} &   1.2\phantom{} &   1.2\phantom{} &   1.2\phantom{} &   1.2\phantom{} \\
\hline
\multirow{3}{*}{3000}
&       $e$ &      1.3\phantom{} &   1.4\phantom{} &   1.4\phantom{} &   1.4\phantom{} & \\
&       $\mu$ &    3.4\phantom{} &   3.4\phantom{} &   3.4\phantom{} &   3.4\phantom{} & \\
&       $e\mu$ &   1.1\phantom{} &   1.1\phantom{} &   1.1\phantom{} &   1.1\phantom{} &   1.1\phantom{} \\
\hline
\hline
\end{tabular}
\end{center}
\end{table}

\begin{table}[!htbp]
\caption{\wp\ and \wstar\ mass limits for the electron and muon decay channels
         and their combination.
         The first column is the decay channel and
         the following give the expected (Exp.) and observed (Obs.) mass limits
         for the SSM \wp\ and for the \wstar\ with equivalent couplings (i.e.\ chosen to
         produce the same decay width as the SSM \wp).
         Masses below the reported limit are excluded by this search.
\label{tab:limits_mass}
}
\begin{center}
\begin{tabular}{c|rr|rr}
\hline
\hline
        &  \multicolumn{4}{c}{Mass limit [\tev]} \\
        &  \multicolumn{2}{c|}{\wp} & \multicolumn{2}{c}{\wstar} \\
        &  Exp. & Obs. &  Exp. & Obs. \\
\hline
 $e$    & 2.50 & 2.50 & 2.38 & 2.38 \\
 $\mu$  & 2.38 & 2.28 & 2.25 & 2.09 \\
 $e\mu$ & 2.55 & 2.55 & 2.42 & 2.42 \\
\hline
\hline
\end{tabular}
\end{center}
\end{table}

The limits presented here are a significant improvement over those reported in previous ATLAS analyses.
Figure~\ref{fig:sigrat_comparison} shows the new and previous ATLAS \xbr\ limits for \wpl\
along with the most recent results from CMS~\cite{cms:wprime2011-2} and CDF~\cite{cdf:Wprime2010}.
Compared with the previous ATLAS results, the limits presented here cover a wider mass range and are about
a factor of five lower at the upper end of the range where they overlap.
Limits from CMS based on data from the same LHC run period are similar.

\begin{figure}[!tbp]
  \centering
  \includegraphics[width=0.49\textwidth]{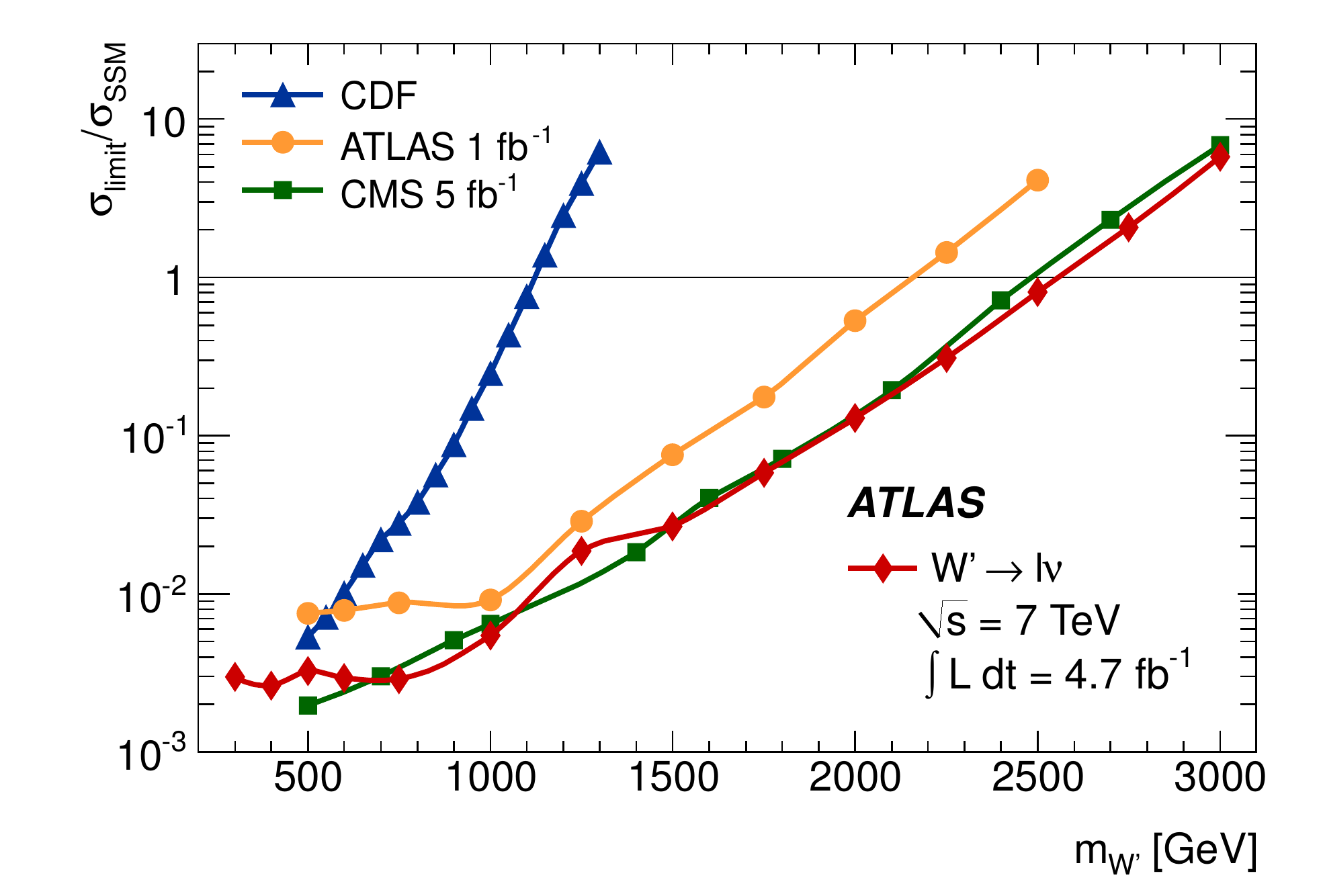}
  \caption{
  Normalised cross-section limits ($\sigma_{\rm limit}/\sigma_{\rm SSM}$) for \wpl\
  as a function of mass for this measurement and from CDF, CMS and the previous ATLAS search.
  The cross-section
  calculations assume the \wp\ has the same couplings as the SM \w~boson.
  The region above each curve is excluded at the 95\% CL.
  \label{fig:sigrat_comparison}
  }
\end{figure}

\section{Conclusions}

The ATLAS detector has been used to search for new high-mass states decaying to a
lepton plus \mettext\ in \pp\ collisions at $\sqrt{s}=7\tev$ using 4.7~\ifb\ of integrated luminosity.
No excess beyond SM expectations is observed.
Bayesian limits on \xbr\ are shown in Figs.~\ref{fig:limits_xbr} and~\ref{fig:sigrat_comparison}.
A \wp\ with SSM couplings is excluded for $\mwp < 2.55~\tev$ at the 95\% CL
and a \wstar\ with equivalent couplings for $\mws < 2.42\tev$.

\begin{acknowledgements}

We thank CERN for the very successful operation of the LHC, as well as the
support staff from our institutions without whom ATLAS could not be
operated efficiently.

We acknowledge the support of ANPCyT, Argentina; YerPhI, Armenia; ARC,
Australia; BMWF, Austria; ANAS, Azerbaijan; SSTC, Belarus; CNPq and FAPESP,
Brazil; NSERC, NRC and CFI, Canada; CERN; CONICYT, Chile; CAS, MOST and NSFC,
China; COLCIENCIAS, Colombia; MSMT CR, MPO CR and VSC CR, Czech Republic;
DNRF, DNSRC and Lundbeck Foundation, Denmark; EPLANET and ERC, European Union;
IN2P3-CNRS, CEA-DSM/IRFU, France; GNAS, Georgia; BMBF, DFG, HGF, MPG and AvH
Foundation, Germany; GSRT, Greece; ISF, MINERVA, GIF, DIP and Benoziyo Center,
Israel; INFN, Italy; MEXT and JSPS, Japan; CNRST, Morocco; FOM and NWO,
Netherlands; RCN, Norway; MNiSW, Poland; GRICES and FCT, Portugal; MERYS
(MECTS), Romania; MES of Russia and ROSATOM, Russian Federation; JINR; MSTD,
Serbia; MSSR, Slovakia; ARRS and MVZT, Slovenia; DST/NRF, South Africa;
MICINN, Spain; SRC and Wallenberg Foundation, Sweden; SER, SNSF and Cantons of
Bern and Geneva, Switzerland; NSC, Taiwan; TAEK, Turkey; STFC, the Royal
Society and Leverhulme Trust, United Kingdom; DOE and NSF, United States of
America.

The crucial computing support from all WLCG partners is acknowledged
gratefully, in particular from CERN and the ATLAS Tier-1 facilities at
TRIUMF (Canada), NDGF (Denmark, Norway, Sweden), CC-IN2P3 (France),
KIT/GridKA (Germany), INFN-CNAF (Italy), NL-T1 (Netherlands), PIC (Spain),
ASGC (Taiwan), RAL (UK) and BNL (USA) and in the Tier-2 facilities
worldwide.
\end{acknowledgements}

\clearpage

\bibliographystyle{support/epjc_atlas.bst}
\bibliography{support/references}{}

\clearpage
\onecolumn
\input{atlas_authlist}

\end{document}

%% file: atlas_authlist.tex
\begin{flushleft}
{\Large The ATLAS Collaboration}

\bigskip

G.~Aad$^{\rm 48}$,
T.~Abajyan$^{\rm 21}$,
B.~Abbott$^{\rm 111}$,
J.~Abdallah$^{\rm 12}$,
S.~Abdel~Khalek$^{\rm 115}$,
A.A.~Abdelalim$^{\rm 49}$,
O.~Abdinov$^{\rm 11}$,
R.~Aben$^{\rm 105}$,
B.~Abi$^{\rm 112}$,
M.~Abolins$^{\rm 88}$
% ...
\bigskip

% ...
$^{11}$ Institute of Physics, Azerbaijan Academy of Sciences, Baku, Azerbaijan\\
$^{12}$ Institut de F\'{i}sica d'Altes Energies and Departament de F\'{i}sica de la Universitat Aut\`{o}noma de Barcelona and ICREA, Barcelona, Spain\\
% ...
$^{21}$ Physikalisches Institut, University of Bonn, Bonn, Germany\\
% ...
$^{48}$ Fakult\"{a}t f\"{u}r Mathematik und Physik, Albert-Ludwigs-Universit\"{a}t, Freiburg, Germany\\
$^{49}$ Section de Physique, Universit\'{e} de Gen\`{e}ve, Geneva, Switzerland\\
% ...
$^{88}$ Department of Physics and Astronomy, Michigan State University, East Lansing MI, United States of America\\
% ...
$^{105}$ Nikhef National Institute for Subatomic Physics and University of Amsterdam, Amsterdam, Netherlands\\
% ...
$^{111}$ Homer L. Dodge Department of Physics and Astronomy, University of Oklahoma, Norman OK, United States of America\\
$^{112}$ Department of Physics, Oklahoma State University, Stillwater OK, United States of America\\
% ...
$^{115}$ LAL, Universit\'{e} Paris-Sud and CNRS/IN2P3, Orsay, France\\
% ...
\end{flushleft}
